\documentclass[12pt]{article}   
   
\usepackage{array}   
\usepackage{epsfig}   
\usepackage{amssymb}   
\usepackage{graphics,graphpap}   
\usepackage{amssymb}   
\usepackage{amsmath}   
\usepackage[usenames]{color}  
\usepackage{slashed}   
\usepackage{graphicx}				  
   
\renewcommand{\thefootnote}{\fnsymbol{footnote}}   
\def \cez {{c_{12}}}   
\def \ced {{c_{13}}}   
\def \cev {{c_{14}}}   
\def \czd {{c_{23}}}   
\def \czv {{c_{24}}}   
\def \cdv {{c_{34}}}   
\def \sez {{s_{12}}}   
\def \sed {{s_{13}}}   
\def \sev {{s_{14}}}   
\def \szd {{s_{23}}}   
\def \szv {{s_{24}}}   
\def \sdv {{s_{34}}}   
\def \ded {{\delta_{13}}}   
\def \dev {{\delta_{14}}}   
\def \dzv {{\delta_{24}}}   
\def \dzv {{\delta_{24}}}   
   
\newcommand{\nn}{\nonumber}

\def\s#1{\setbox0=\hbox{$#1$}%
  \rlap{\ifdim\wd0>.7em\kern.22\wd0\else\kern.1\wd0\fi /}#1}

   
\begin{document}   
   
\begin{titlepage}   
\begin{flushright}   
\end{flushright}   
\vskip1.5cm   
\begin{center}   
{\Large \bf \boldmath   
How much space is left for a new family?}   
\vskip1.3cm    
{\sc   
Markus Bobrowski    \footnote{Markus.Bobrowski@physik.uni-regensburg.de}$^{,1}$,   
Alexander Lenz      \footnote{Alexander.Lenz@physik.uni-regensburg.de}$^{,1}$,   
Johann Riedl        \footnote{Johann.Riedl@physik.uni-regensburg.de}$^{,1}$   
\\ and \\   
J{\"u}rgen Rohrwild \footnote{Juergen.Rohrwild@physik.uni-regensburg.de}$^{,1}$}   
\vskip0.5cm   
$^1$ Institut f\"ur Theoretische Physik, \\ Universit\"at Regensburg,   
D--93040 Regensburg, Germany   
   
\vskip2cm

   
\vskip3cm   
   
{\large\bf Abstract\\[10pt]} \parbox[t]{\textwidth}{   
We perform an exploratory study of the allowed parameter range for the CKM-like   
mixing of hypothetical quarks of a fourth generation. As experimental constraints    
we use the tree-level determinations of the 3$\times$3 CKM elements and    
FCNC processes ($K$-, $D$-, $B_d$-, $B_s$-mixing and the decay $b \to s \gamma$)   
under the assumption that the 4$\times$4 CKM matrix is unitary. For the FCNCs  
we use some simplifying assumptions concerning the QCD corrections.   
Typically small mixing with the fourth family is favoured; contrary to    
expectation, however, we find that also a quite large mixing with the 4th family    
is not yet excluded.   
}   
   
\vfill   
   
\end{center}   
\end{titlepage}   
   
\setcounter{footnote}{0}   
\renewcommand{\thefootnote}{\arabic{footnote}}   
\renewcommand{\theequation}{\arabic{section}.\arabic{equation}}   
   
\newpage   
   
\section{Introduction}   
\setcounter{equation}{0}   
Additional particle generations have been discarded for a long time.   
Recently this possibility (see \cite{Frampton:1999xi} for a review) gained more interest. 
In contrast to many previous claims a fourth family is not in conflict with electroweak 
precision tests \cite{Kribs:2007nz}, see also \cite{Holdom:1996bn,He:2001tp,Novikov:2002tk}
 for earlier works.  
The authors of \cite{Kribs:2007nz} have shown that if the quark masses of the 4th generation   
fulfill the following relation  
\begin{equation}  
m_{t'} - m_{b'} \approx \left( 1+ \frac15 \ln \frac{m_H}{115 \, \mbox{GeV}} \right) \times 55 \, \mbox{GeV} , 
\end{equation}  
the electro-weak oblique parameters \cite{Peskin:1991sw} are within the   
experimentally allowed regions.  
This also has the crucial side effect that a fourth generation softens the current Higgs bounds,  
see e.g. \cite{Flacher:2008zq}.  
Moreover, an additional family might solve problems related to baryogenesis.   
First, it could lead to a sizeable increase of the measure of CP-violation, see \cite{Hou:2008xd}. 
Second it also would increase the strength of the phase transition, see \cite{Carena:2004ha}.   
In addition,  the gauge couplings  can be unified
without invoking SUSY \cite{Hung:1997zj}.   
A new family also might cure certain problems in flavor physics, see e.g.   
\cite{Soni:2008bc,Hou:2006mx,Arhrib:2006pm,Hou:2005hd} for some recent work and  
e.g. \cite{Hou:1986ug,Hou:1987vd} for some early work on 4th generation effects on  
flavor physics. 
\\  
In view of the (re)start of the LHC,  it is important not to exclude any  possibility  
for new physics scenarios simply due to prejudices.  
\\  
In this work we, therefore, perform an exploratory study of the allowed parameter range  
for the CKM-like mixing of hypothetical quarks of a fourth generation.  
In Section 2 we first describe the general parameterization used for the four generation CKM matrix, 
next we explain the  experimental constraints for the quark mixing. We then describe the 
numerical scan through the parameter space and finally we present  the allowed parameter ranges 
for the mixing with an additional family. In Section 3 we perform a Taylor expansion of the 
4$\times$4   
CKM matrix \`{a} la Wolfenstein, which makes the complicated general parameterization of $V_{CKM4}$  
much clearer; in particular the possible hierarchy of the mixing is  clearly visible.  
In Section 4 we discuss some peculiar parameter ranges, which show huge deviations  
from  current knowledge of the threedimensional CKM matrix, and  explain why these effects 
are not seen in the current CKM fits. Finally we conclude with an outlook on possible extensions 
of this exploratory study.  
\section{Constraints on $V_{CKM4}$}   
\setcounter{equation}{0}   
\subsection{Parameterization of $V_{CKM4}$}   
Let the minimal standard model with three generations of fermions be denoted by SM3. The mixing  
between  quarks is described by the unitary 3 dimensional CKM-matrix \cite{Cabibbo:1963yz, Kobayashi:1973fv},  
which can be parameterized by three angles, $\theta_{12}$,  $\theta_{13}$ and $\theta_{23}$  
($\theta_{ij}$ describes the strength of the mixing between the $i$th and $j$th family) and  
the CP-violating phase  $\delta_{13}$. The so-called standard parameterization of $V_{CKM3}$   
reads   
\begin{equation}  
                   V_{CKM3} =   
\left( \begin{array}{ccc} c_{12} c_{13}  & s_{12} c_{13} & s_{13} e^{-i \delta_{13}} \\   
  -s_{12} c_{23} - c_{12} s_{23} s_{13} e^{i \delta_{13}} &  c_{12} c_{23} - s_{12} s_{23} s_{13} e^{i \delta_{13}} & s_{23} c_{13} \\  
   s_{12} s_{23} - c_{12} c_{23} s_{13} e^{i \delta_{13}} & -c_{12} s_{23} - s_{12} c_{23} s_{13} e^{i \delta_{13}} & c_{23} c_{13} \\  
       \end{array}   
\right)  
                   \end{equation}  
with  
\begin{equation}  
s_{ij} := \sin (\theta_{ij}) \, \, \, \, \, \mbox{and} \, \, \, \, \,  
c_{ij} := \cos (\theta_{ij})\, .  
\end{equation}  
Extending the minimal standard model to include a fourth family of fermions (SM4) introduces at least    
14 new parameters. We do not take into account any correlations to the mixing matrix of the leptons. The seven parameters that are directly related to the  quark sector
\begin{itemize}   
\item 3 additional angles in the CKM-matrix, which we denote by $\theta_{14}, \theta_{24}$   
      and $\theta_{34}$,   
\item 2 additional CP-violating phases in the CKM-matrix: $\delta_{14}$ and $\delta_{24}$,   
\item 2 quark masses of the 4th family: $m_{b'}$ and $m_{t'}.$   
\end{itemize}   
For the quark masses we have  bounds from direct searches at TeVatron   
\cite{Aaltonen:2007je,:2008nf}  
\begin{equation}   
m_{b'}  >  268   \, \mbox{GeV}, \, \, \, \, \, \, \,    
m_{t'}  >  256   \, \mbox{GeV}.     
\end{equation}   
In \cite{Hung:2007ak} it was claimed that in deriving these bounds implicit assumptions    
about the couplings of the fourth family have been made. Without these assumptions the mass bounds can be weaker.  
We investigate the following mass parameter range - taking into account the results of  
\cite{Kribs:2007nz}   
\begin{eqnarray}  
\label{eq:masstp}300 \, \mbox{GeV} \leq m_{t'} & \leq & 650  \, \mbox{GeV} \, ,  
\\   
 m_{b'} & = & m_{t'} -   55 \, \mbox{GeV} \, ,  
\\  
245 \, \mbox{GeV} \leq m_{b'} & \leq & 595  \, \mbox{GeV} \, .  
\end{eqnarray}  
Our goal is the determination of the current experimentally allowed ranges for the parameters   
$\theta_{14}$, $\theta_{24}$, $\theta_{34}$,  $\delta_{14}$ and $\delta_{24}$.   
For our numerical analysis we use an exact parameterization of the four-dimensional CKM matrix.    
The form suggested by Fritzsch and Plankl \cite{Fritzsch:1986gv}\footnote{In the original paper    
of Fritzsch and Plankl there is a typo in the element $V_{cb}$: $c_{23}$ has to be replaced by $s_{23}$.}   
and simultaneously by Harari and Leurer \cite{Harari:1986xf} turns out to be especially useful,   
because in the limiting case of vanishing mixing  with the fourth family the standard   
parameterization of the 3$\times$3 CKM matrix is restored. Moreover, this form of the matrix reveals a particularly convenient structure: the simplicity of the first row 
is advantageous because these elements are experimentally very well constrained, while the 
compact form of the last column simplifies the Taylor expansion presented in Section 3.  
  
{\footnotesize   
\begin{equation}   
 V_{CKM}^{(4)} = \left( \begin{array}{cccc}   
\cez \ced \cev  & \ced \cev \sez & \cev \sed e^{-i\ded} & \sev e^{-i\dev} \\   
& & &\\   
-\czd \czv \sez -\cez \czv \sed \szd e^{i\ded} &\cez \czd \czv -\czv \sez \sed \szd e^{i\ded} &   
      \ced \czv \szd & \cev \szv e^{-i\dzv} \\   
-\cez \ced \sev \szv e^{i(\dev-\dzv)} & -\ced \sez \sev \szv e^{i(\dev-\dzv)} &   
      -\sed \sev \szv e^{-i(\ded+\dzv-\dev)} &  \\   
& & & \\   
-\cez \czd \cdv\sed e^{i\ded} +\cdv \sez \szd & -\cez \cdv \szd - \czd \cdv \sez \sed e^{i\ded} &   
      \ced\czd\cdv & \cev \czv \sdv  \\   
-\cez \ced \czv \sev \sdv e^{i \dev} & -\cez \czd \szv\sdv e^{i\dzv} & -\ced\szd\szv\sdv e^{i\dzv} &   
 \\   
+\czd\sez\szv\sdv e^{i\dzv}& -\ced \czv \sez\sev\sdv e^{i\dev} & -\czv\sed\sev\sdv e^{i(\dev-\ded)} &   
      \\   
+\cez \sed \szd\szv\sdv e^{i(\ded+\dzv)} & +\sez \sed \szd \szv \sdv e^{i(\ded+\dzv)} & & \\   
& & &\\   
-\cez \ced \czv \cdv \sev e^{i\dev } & -\cez \czd \cdv \szv e^{i\dzv}+\cez \szd \sdv &   
       -\ced \czd \sdv & \cev \czv \cdv\\   
+\cez \czd \sed \sdv e^{i \ded} & -\ced \czv \cdv \sez \sev e^{i \dev}&   
       -\ced \cdv \szd \szv e^{i\dzv} & \\   
+\czd \cdv \sez \szv e^{i\dzv}-\sez \szd \sdv & +\czd \sez \sed \sdv e^{i \ded} &   
       -\czv \cdv \sed \sev e^{i(\dev-\ded)} & \\   
+\cez \cdv \sed \szd \szv e^{i(\ded+\dzv)} & +\cdv \sez  \sed \szd \szv e^{i(\ded + \dzv)} & &   
\end{array} \right) \label{eq:CKM4FP}  
\end{equation}   
}   
\subsection{Experimental bounds}  
In this section we summarize the experimental constraints that have to be 
fulfilled by the quark mixing matrix. 
The elements of the 3$\times$3 CKM matrix have been studied   
intensely for many years and precision data on most of them   
is available. In principle there are two different ways to  
determine the matrix elements. On the one hand, they 
enter charged weak decays already at tree-level and a measurement  
of e.g. the corresponding decay rate provides direct information on  
the  CKM elements (see e.g. \cite{Battaglia:2003in} and references therein).  
We will refer to such constraints as {\it   
tree-level constraints}. On the other hand, processes involving  
a flavor-changing neutral current (FCNC) are forbidden at tree-level  
and only come into play at loop level via the renowned Penguin and  
Box diagrams. These processes provide strong bounds,   
 referred to as {\it FCNC constraints}, on the structure of the   
CKM matrix and its elements.   
In what follows we discuss the implications of these constraints  
in more detail.   
\paragraph{Tree-level constraints for the CKM parameters:}  
Since the (absolute) value of only one CKM element enters the   
theoretical predictions for weak tree-level decays, no GIM mechanism  
or unitary condition has to be assumed. By matching theory and   
experiment the matrix element can be extracted {\it independently} of the   
number of generations. Therefore, all tree-level constraints have   
the same impact on the 4$\times$4 matrix as they have on the    
3$\times$3 one.  
   
We take the PDG values \cite{Amsler:2008zzb} for our analysis:  
 \begin{center}  
  \begin{tabular}{|c||c|c|c|}  
\hline  
   & absolute value  & relative error & direct measurement from \\ \hline  
$V_{ud}$ & $0.97418 \pm 0.00027$ & $0.028\%$& nuclear beta decay \\\hline  
$V_{us}$ & $0.2255  \pm 0.0019$  & $0.84\%$ & semi-leptonic K-decay\\\hline  
$V_{ub}$ & $0.00393 \pm 0.00036$ & $9.2\%$  & semi-leptonic B-decay\\\hline  
$V_{cd}$ & $0.230   \pm 0.011$   & $4.8\%$  & semi-leptonic D-decay\\\hline  
$V_{cs}$ & $1.04    \pm 0.06$    & $5.8\%$  & (semi-)leptonic D-decay\\\hline  
$V_{cb}$ & $0.0412  \pm 0.0011$  & $2.7\%$  & semi-leptonic B-decay\\\hline  
$V_{tb}$ & $>0.74$               &          & (single) top-production\\\hline  
  \end{tabular}  
 \end{center}  
In the following, we denote the absolute values in the table above as $|V_i|\pm\Delta V_i$.  
Next, we will discuss the bounds coming from  FCNCs.  
\paragraph{FCNC constraints:}  
      It is well known that FCNC processes give strong constraints on extensions of the    
      standard model. In particular information about the CKM elements $V_{tx}$ can be obtained   
      by investigating $B$-and $K$-mixing.   
      The mixing of the neutral mesons is described by box diagrams. As an example we show   
      the box diagrams for $B_d$ mixing:  
      \begin{center}  
      \includegraphics[width=0.9\linewidth]{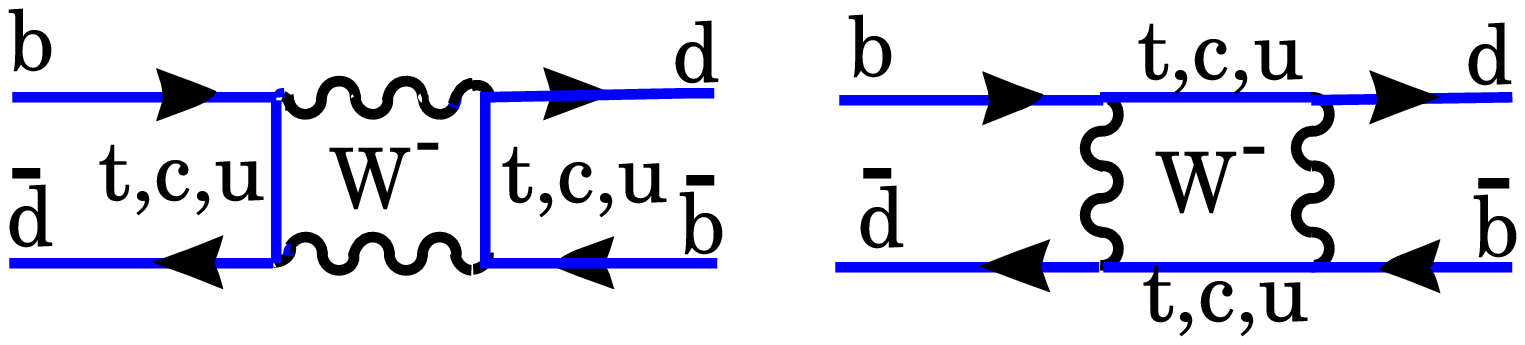}   
      \end{center}  
      $M_{12}$ encodes the   
      virtual part of the box diagrams, which is very sensitive to new physics contributions. It is    
      related to the mass difference of the neutral mesons via   
      \begin{equation}   
      \Delta M = 2 | M_{12}|\, .   
      \end{equation}   
      In the SM3 one obtains the following relations  
      \begin{eqnarray}   
      M_{12}^{K^0} & \propto & \eta_{cc} \left(\lambda_c^{K^0}\right)^2 S_0(x_c)    
                         + 2 \eta_{ct} \lambda_c^{K^0} \lambda_t^{K^0} S(x_c,x_t)    
                         + \eta_{tt} \left(\lambda_t^{K^0}\right)^2 S_0(x_t) \,  ,  
      \\   
      M_{12}^{B_d} & \propto & \eta_{tt} \left(\lambda_t^{B_d}\right)^2 S_0(x_t) \, ,  
      \\   
      M_{12}^{B_s} & \propto & \eta_{tt} \left(\lambda_t^{B_s}\right)^2 S_0(x_t) \, ,  
        \end{eqnarray}   
      with the Inami-Lim functions \cite{Inami:1980fz}  
     \begin{eqnarray}   
     S_0(x) & = & \frac{4x - 11 x^2 + x^3}{4(1-x)^2} - \frac{3 x^3 \ln[x]}{2 (1-x)^3} \, ,  
     \\   
     S(x,y) & = & x y  \left[   
                  \frac{1}{y-x} \left( \frac14 + \frac32 \frac{1}{1-y} - \frac34 \frac{1}{(1-y)^2} \right) \ln[y]   
        \right. \nonumber \\    
          &&  \left.   + \frac{1}{x-y} \left( \frac14 + \frac32 \frac{1}{1-x} - \frac34 \frac{1}{(1-x)^2} \right) \ln[x]   
              - \frac34  \frac{1}{1-x} \frac{1}{1-y} \right] \, ,  
     \end{eqnarray}   
     where $ x_t  =  \frac{m_t^2}{M_W^2}$,  the CKM elements   
     \begin{equation}   
     \lambda_{x}^{K^0} = V_{xd} V_{xs}^*,  \, \, \, \, \, \, \,     
     \lambda_{x}^{B_d} = V_{xd} V_{xb}^*,  \, \, \, \, \, \, \,     
     \lambda_{x}^{B_s} = V_{xs} V_{xb}^*       
     \end{equation}   
     and the QCD corrections \cite{Buras:1990fn,Herrlich:1993yv,Herrlich:1996vf}  
     \begin{equation}   
     \eta_{cc}= 1.38   \pm 0.3    ,  \, \, \, \, \, \, \,    
     \eta_{ct}= 0.47   \pm 0.04   ,  \, \, \, \, \, \, \,    
     \eta_{tt}= 0.5765 \pm 0.0065 .   
     \end{equation}   
     The full expressions for $M_{12}$ can be found e.g. in \cite{Buras:1990fn,Lenz:2006hd}.   
     In deriving these expressions unitarity (of the $3\times 3$ matrix) was explicitly used, i.e.  
     \begin{equation}   
     \lambda_u^X + \lambda_c^X + \lambda_t^X = 0 \, .   
     \end{equation}   
     Moreover, in the $B$-system the CKM-elements of the different internal quark    
     contributions are all roughly of the same size. Only the top contribution, which has by far the largest   
     value of the Inami-Lim functions, survives. This is not the case in the $K$-system. Here   
     the top contribution is CKM suppressed, while the kinematically suppressed charm terms are   
     CKM favored. Therefore, both have to be taken into account.   
     More information about the mixing of neutral mesons can be found e.g. in    
     \cite{Lenz:2006hd, Anikeev:2001rk}.  
     \\   
     For the mixing of neutral mesons we define the parameter $\Delta$    
     that quantifies the deviation from the standard model \cite{Lenz:2006hd}:    
      \begin{equation}   
      \Delta = \frac{M_{12}^{SM4}}{M_{12}^{SM3}} = |\Delta| e^{i \phi^{\Delta}}\, .   
      \end{equation}   
    Going over to the SM4, we obtain   
      \begin{eqnarray}   
      M_{12}^{K^0, SM4} & \propto & \eta_{cc} \left(\lambda_c^{K^0}\right)^2 S_0(x_c)    
                         + 2 \eta_{ct} \lambda_c^{K^0} \lambda_t^{K^0} S(x_c,x_t)    
                         + \eta_{tt} \left(\lambda_t^{K^0}\right)^2 S_0(x_t)   
      \\ &&   
                         + 2 \eta_{ct'} \lambda_t^{K^0} \lambda_{t'}^{K^0} S(x_c,x_{t'})    
                         + 2 \eta_{tt'} \lambda_t^{K^0} \lambda_{t'}^{K^0} S(x_t,x_{t'})   
                         + \eta_{t't'}  \left(\lambda_{t'}^{K^0}\right)^2 S_0(x_{t'}) \, ,        
       \nonumber \\   
      M_{12}^{B_d, SM4} & \propto & \eta_{tt}   \left(\lambda_t^{B_d}\right)^2 S_0(x_t)    
                                   +\eta_{t't'} \left(\lambda_{t'}^{B_d}\right)^2 S_0(x_{t'})   
                                   + 2 \eta_{tt'} \lambda_t^{B_d} \lambda_{t'}^{B_d} S(x_t,x_{t'})  \, ,  
      \\   
      M_{12}^{B_s, SM4} & \propto & \eta_{tt} \left(\lambda_t^{B_s}\right)^2 S_0(x_t)   
                                   +\eta_{t't'} \left(\lambda_{t'}^{B_s}\right)^2 S_0(x_{t'})   
                                   + 2 \eta_{tt'} \lambda_t^{B_s} \lambda_{t'}^{B_s} S(x_t,x_{t'}) \, .  
        \end{eqnarray}   
     Note that now also those CKM elements change that describe the mixing within the first three families!  
     For simplicity we take the new QCD corrections to be  
     \begin{equation}  
     \eta_{t't'} = \eta_{tt'} = \eta_{tt} \, \, \, \mbox{and} \, \, \,  \eta_{ct'} =   \eta_{ct} \, .  
     \end{equation}   
     In addition to the mixing quantities we also investigate the decay $b \to s \gamma$. To obtain the  
     SM4 prediction for $b \to s \gamma$ one has to do the whole analysis of this decay without invoking  
     the unitarity of the 3$\times$3 CKM matrix, which is beyond the scope of this work. As an estimate  
     of the effects of a fourth generation on $b \to s \gamma$, we simply define the ratio of the CKM structure  
     times the corresponding Inami-Lim function $D_0'(x_t)$ \cite{Inami:1980fz}\footnote{The Inami-Lim   
     function $D_0'(x_t)$ is proportional to the Wilson-coefficient $C_{7\gamma}(M_W)$.}:  
     \begin{equation}   
     \Delta_{b \to s \gamma} :=   
      \frac{|\lambda_t^{SM4}|^2 D_0'(x_t)^2 + 2 \mbox{Re} \left(\lambda_t^{SM4} \lambda_{t'}^{SM4} \right) D_0'(x_t) D_0'(x_{t'})   
            +|\lambda_{t'}^{SM4}|^2 D_0'(x_{t'})^2}  
            {|\lambda_t^{SM3}|^2 D_0'(x_t)^2} \, ,  
     \end{equation}   
     with  
     \begin{equation}  
     D_0'(x) = -\frac{-7x +  5x^2+8x^3}{12(1-x)^3}   
                 + \frac{x^2 (2-3x)}{2(1-x)^4} \ln[x] \, .  
     \end{equation}  
     Parameters which give a value of $\Delta_{b \to s \gamma}$ close to one will also lead only to small  
     deviations of $\Gamma (b \to s \gamma)^{SM4}/ \Gamma (b \to s \gamma)^{SM3}$ from one.

\begin{figure}   
\centering
\begin{minipage}{0.85\textwidth}   
\begin{center} 
\includegraphics[width=0.8\linewidth]{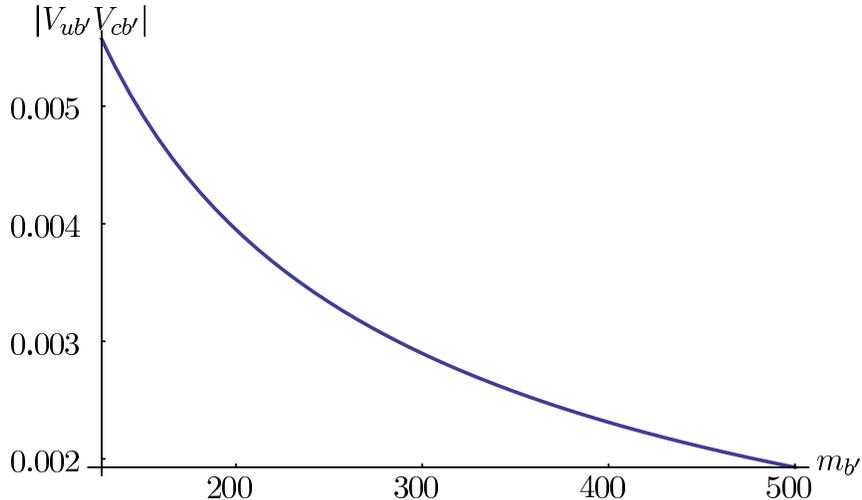}   
\caption{Bound on $|V_{ub'} V_{cb'}|$ determined from the measurement of $D^0$-mixing in dependence   
         on the mass of the $b'$ quark. \label{Dmixbound}}   
\end{center} 
\end{minipage}   
\end{figure}

     Currently, in particular the hadronic uncertainties are under intense discussion, see e.g.   
     \cite{Lenz:2008xt}. Therefore, we use two sets of bounds for the allowed deviations from the 
      SM3 values, which cover the possible range of  uncertainties, a {\it conservative} and an {\it aggressive} one:  
      \begin{displaymath}   
      \begin{array}{|c|c|c|}   
      \hline   
                          &   \mbox{Conservative Bound} & \mbox{Aggressive Bound}   
      \\   
      \hline \hline   
      |\Delta_{B_d}|      & 1 \pm 0.3      & 1 \pm 0.1   
      \\   
      \hline   
      \phi^\Delta_{B_d}   & 0 \pm 10^\circ      & 0 \pm 5^\circ   
      \\   
      \hline   
      |\Delta_{B_s}|      & 1 \pm 0.3      & 1 \pm 0.1   
      \\   
      \hline   
      \phi^\Delta_{B_s}   & \mbox{free}    & \mbox{free}   
      \\   
      \hline   
      \mbox{Re} (\Delta_{K})      & 1 \pm 0.5      & 1 \pm 0.25   
      \\   
      \hline   
      \mbox{Im} (\Delta_{K})      & 0 \pm 0.3      & 0 \pm 0.15   
      \\   
      \hline   
      \Delta_{b \to s \gamma}      & 1 \pm 0.15      & 1 \pm 0.07   
      \\   
      \hline   
       \end{array}   
      \end{displaymath}   
In  \cite{Golowich:2007ka} a very strong bound on $|V_{ub'} V_{cb'}|$ is    
extracted from $D^0$-mixing. We redo this analysis and confirm the conclusion   
of \cite{Golowich:2007ka}, although we are able to soften the bound by a factor of $\sqrt{3}$.   
The starting point is the mass difference in the neutral $D^0$-system, which can be expressed   
in terms of the parameter $x_D$:   
\begin{equation}   
x_D = \frac{\Delta M_D}{\Gamma_D} = \frac{2 |M_{12}^{D^0}|}{\Gamma_D} \, .   
\end{equation}   
HFAG \cite{Barberio:2001mb} quotes for an experimental value of $x_D$  
\begin{equation}   
x_D = (0.811 \pm 0.334) \cdot 10^{-2} \, .   
\end{equation}   
Starting with the expression for the box diagram and using the unitarity condition   
$\lambda_d^{D^0} + \lambda_s^{D^0} + \lambda_b^{D^0} + \lambda_{b'}^{D^0} = 0$    
(with $\lambda_{x}^{D^0} = V_{cx} V_{ux}^*$ ),   
we obtain   
\begin{eqnarray}\label{09022401mb}  
M_{12}^{D^0} & \propto & \lambda_s^2 S_0(x_s) + 2 \lambda_s \lambda_b S(x_s, x_b) + \lambda_b^2 S_0(x_b) + \text{LD}
\nonumber   
\\   
&& + 2 \lambda_s \lambda_{b'} S(x_s, x_{b'})+ 2 \lambda_b \lambda_{b'} S(x_b, x_{b'}) + \text{LD}   
\nonumber   
\\   
&& + \lambda_{b'}^2 S_0(x_{b'}), \,   
\end{eqnarray}   
where the proportionality constant is   
\begin{equation}  
\frac{{G_{\text{F}}^2 M_W^2 M_D }}  
{{12\pi ^2 }}\;f_D^2 B_D \;\eta \left( {m_c ,M_W } \right).  
\end{equation}  
Lubicz and Tarantino \cite{Lubicz:2008mb} gave a survey of recent lattice data and provided an 
averaged decay constant $f_{D^0}=212\pm 14$ MeV and bag parameter $B=0.85 \pm 0.09 $. In order to 
compare with the results of \cite{Golowich:2007ka}, we use only the LO expression of the QCD correction factor  
$\eta$,
\begin{equation}  
\eta \left( {m_c ,M_W } \right) \equiv \left( {\frac{{\alpha _s^{\left( 4 \right)} \left( {m_b } \right)}}  
{{a_s^{\left( 4 \right)} \left( {m_c } \right)}}} \right)^{\frac{6}  
{{25}}} \left( {\frac{{\alpha _s^{\left( 5 \right)} \left( {M_W } \right)}}  
{{\alpha _s^{\left( 5 \right)} \left( {m_b } \right)}}} \right)^{\frac{6}  
{{23}}}   
\simeq 0.74 \, .  
\end{equation}  
The first line of (\ref{09022401mb}) corresponds to the pure SM3 contribution, the third line is due to contributions   
of the heavy 4th generation and the second line is a term arising when SM3- and $b'$ contributions mix:   
\begin{eqnarray}   
M_{12}^{D^0} & = & M_{12,SM3}^{D^0} +  M_{12, Mix}^{D^0} +  M_{12, b'}^{D^0} \, .   
\end{eqnarray}    
The perturbative short-distance contribution to $ M_{12,SM3}^{D^0}$ is numerically very small.   
The first two terms in the first line of (\ref{09022401mb}) are kinematically suppressed and   the third   
term suffers a Cabibbo suppression caused by a CKM factor of order $\mathcal{O}\left(10^{-8}\right)$,   
such that an OPE-based standard model  calculation  yields values of about $x \approx 4 \cdot 10^{-5}$. The order of magnitude of this result complies with early estimates for $x_D$, which relied merely on perturbation theory calculations and ranged between roughly $10^{-6}$ \cite{Datta:1985mb} and $10^{-4}$ \cite{Cheng:1982mb}. It has often been pointed out that in the case of charmed mesons a substantial enhancement of the mass and   
width differences has possibly to be attributed to long-distance (LD) effects, which cannot be calculated   
perturbatively, see e.g. \cite{Bigi:2001mb,Golowich:1998mb,Buccella:1996mb,Donoghue:1986mb,Wolfenstein:1985mb}.   
The quoted predictions usually rely on exclusive estimates of decay widths; they   
can be considerably increased by nearby resonances. Typical results are in the range of   
$x_D,y_D\simeq 10^{-4}\dots 10^{-3}$, which almost reach the order of magnitude of the experimental values.   
Bigi and Uraltsev \cite{Bigi:2001mb} argue that, albeit the leading  $1/m_c$    
contributions are negligibly small and the validity of duality is very questionable, operators of higher   
dimension might lead to values of $x_D,y_D$ up to $ 5\cdot 10^{-3}$  in the framework of the   
standard OPE techniques, what is already very close to the experimental values.   
\\  
The short-distance terms of the mixed part $ M_{12, Mix}^{D^0}$ are numerically   
at most as large as the short-distance part of  the pure SM3 contribution.   
The $s$-quark term of the mixed part is about twice  the $b$-quark term and it might also be   
 affected by large long-distance effects. For $ M_{12, b'}^{D^0}$ the OPE is   
expected to work perfectly and no sizeable unknown non-perturbative   
effects are likely to appear. Numerically this term can be much larger than the short-distance 
parts of the SM3- and the mixed contribution.   
\\   
The idea of \cite{Golowich:2007ka} was to neglect all terms in $M_{12}^{D^0}$, except $ M_{12, b'}^{D^0}$,   
and to equate this term with the experimental number for $x_D$. Following this strategy we reproduce the   
bounds given in \cite{Golowich:2007ka}.   
We think, however, that it is not completely excluded that there might be large non-perturbative   
contributions to both $ M_{12,SM3}^{D^0}$  and  $M_{12, Mix}^{D^0}$, each of the size of the experimental   
value of $x_D$. This would enhance the possible range for $ M_{12, b'}^{D^0}$ by a factor of up to 3 compared to   
 \cite{Golowich:2007ka}.   
Allowing this possibility we obtain the following, very conservative bounds on $|V_{ub'} V_{cb'}|$, see also   
Fig.(\ref{Dmixbound}):   
\begin{equation}    
|V_{ub'} V_{cb'}| \leq \left\{   
\begin{array}{cc}   
\label{eq:mass}0.00395 & \, \mbox{for} \, \, \, m_{b'} = 200 \, \mbox{GeV} \, ,   
\\   
0.00290 & \, \mbox{for} \, \, \, m_{b'} = 300 \, \mbox{GeV} \, ,   
\\   
0.00193 & \, \mbox{for} \, \, \, m_{b'} = 500 \, \mbox{GeV} \, .   
\end{array}   
\right.  
\end{equation}   
Even as we were able to soften the bound of \cite{Golowich:2007ka} by a factor $\sqrt{3}$, $D^0$-mixing   
is still by far the strongest direct constraint on $|V_{ub'} V_{cb'}|$.   
We take the values of Eq. (\ref{eq:mass}) for our conservative bounds,  
while we take the results of \cite{Golowich:2007ka} as  
the aggressive ones.  
\subsection{Scan through the mixing parameters}  
\begin{figure}[t]  
 \epsfig{figure=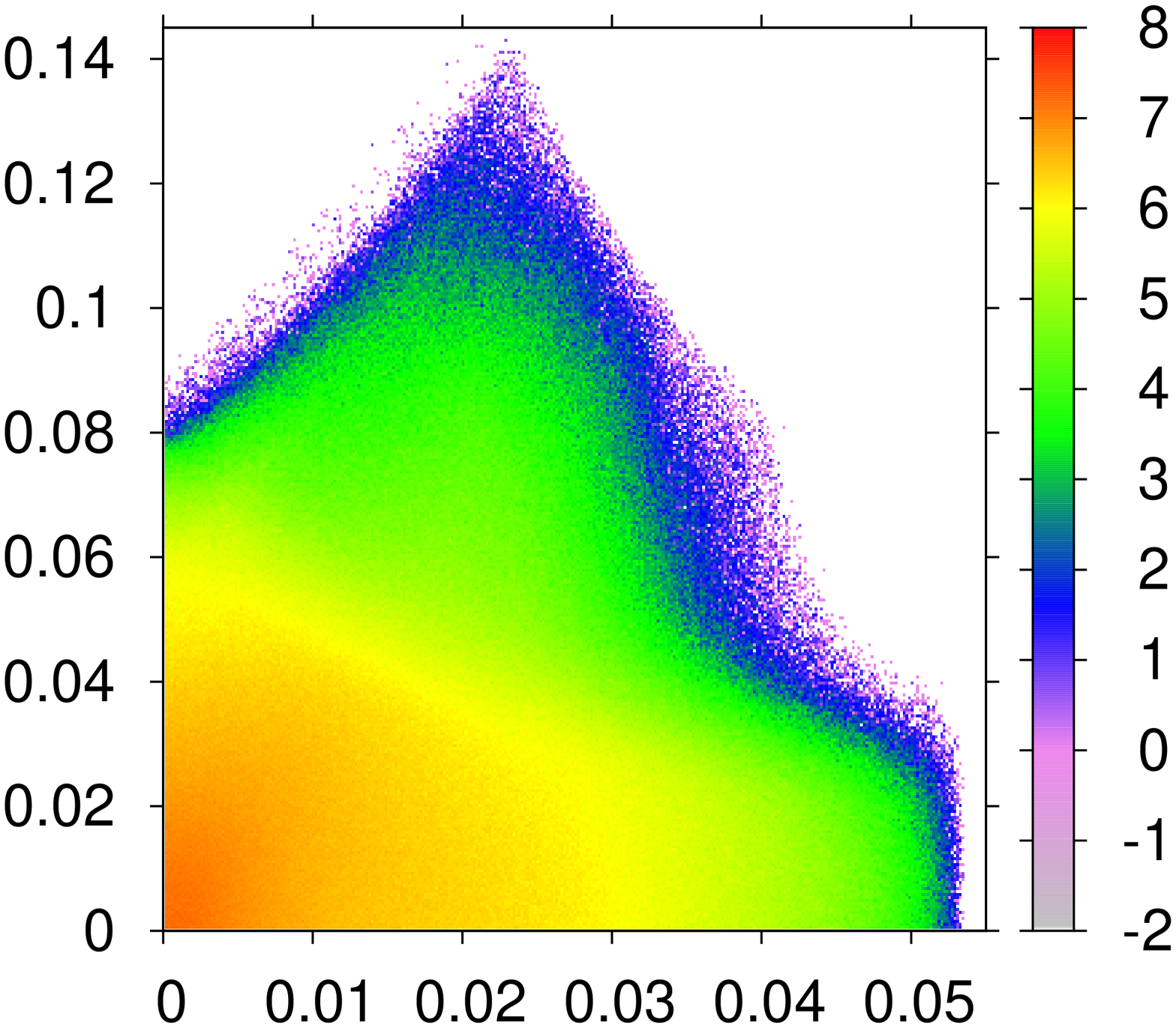,angle=270,width=0.49\textwidth}  
\hfill  
\epsfig{figure=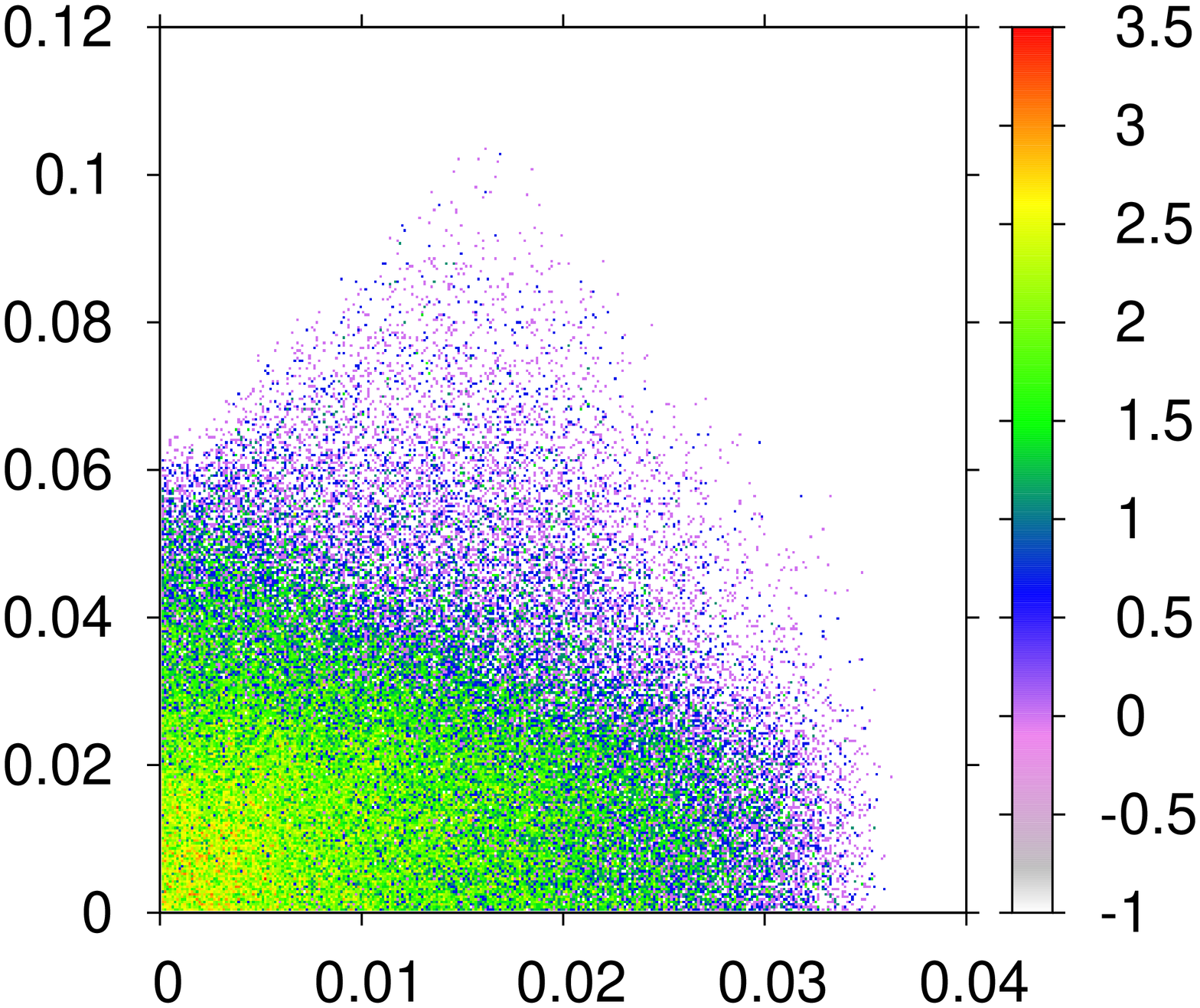,angle=270,width=0.49\textwidth}  
 \epsfig{figure=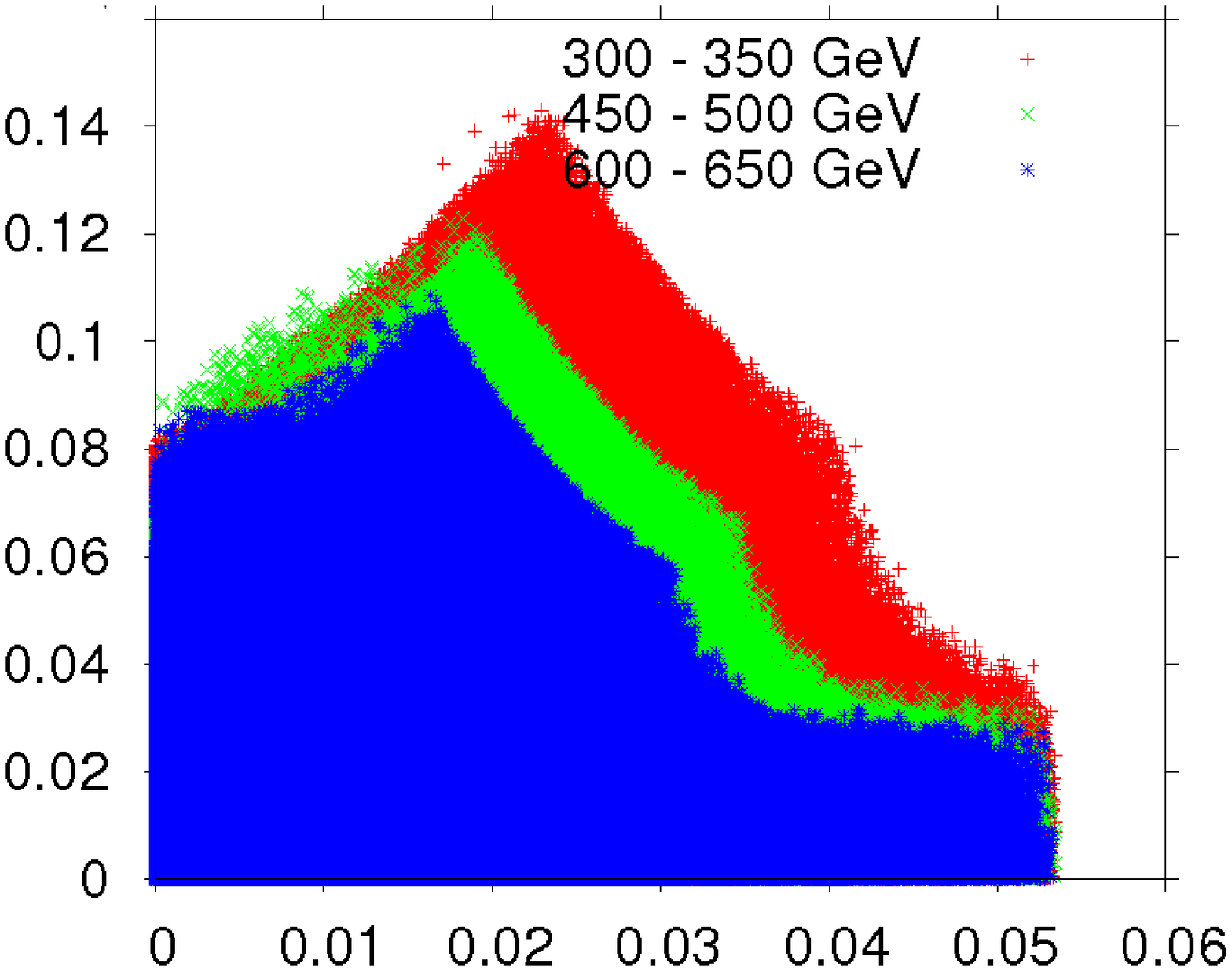,angle=270,width=0.49\textwidth}  
\hfill  
\epsfig{figure=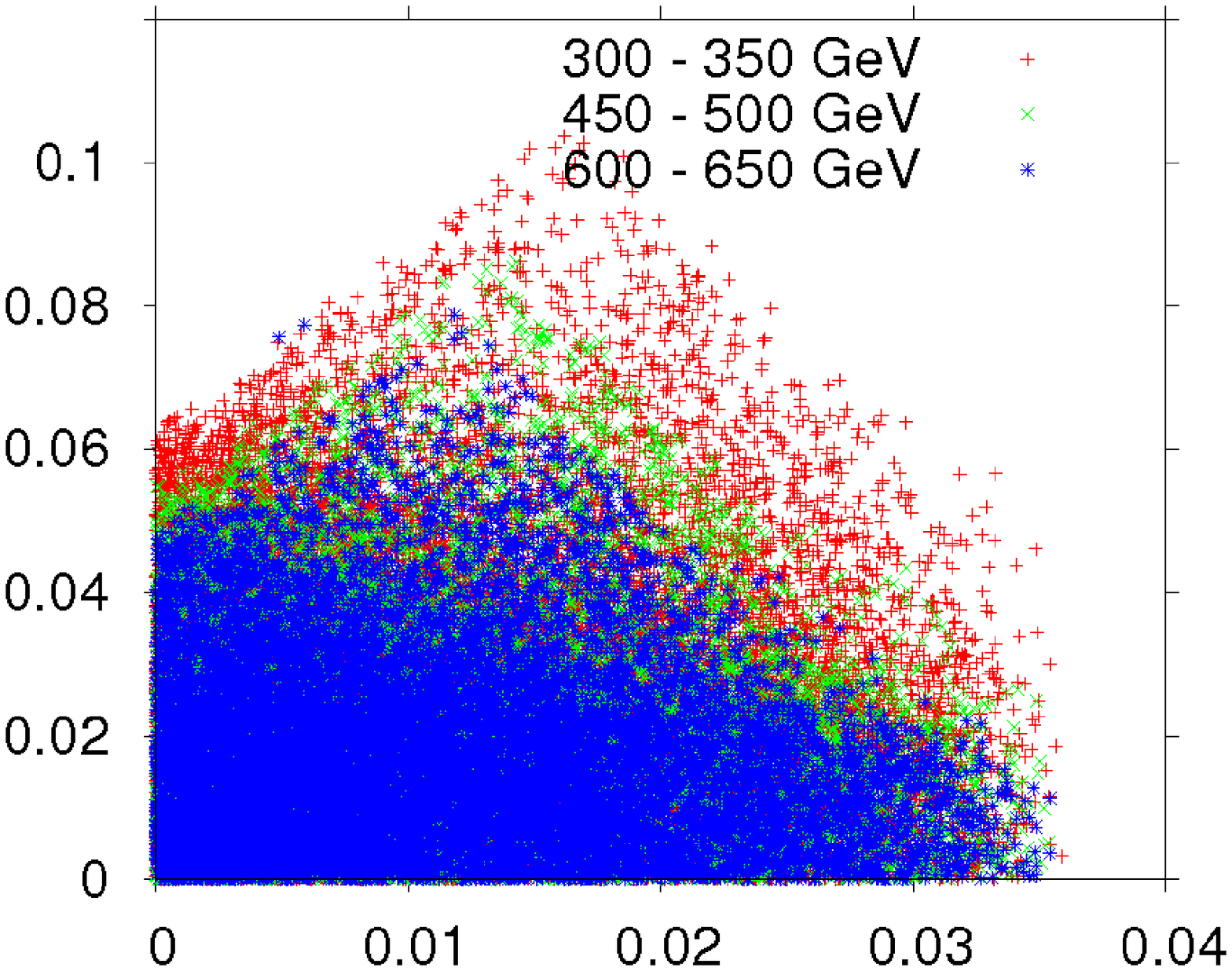,angle=270,width=0.49\textwidth}  
\caption{In the upper left and upper right panel, the allowed parameter ranges  for $\theta_{14}$ on 
the x axis and $\theta_{24}$ on the y axis are shown for the conservative and the aggressive bounds, 
respectively.  The colour encodes the relative occurrence as explained in the text. In the lower 
left and right panels the allowed parameter range is shown in dependence on the $t'$ mass for 
three different mass ranges for the conservative and aggressive bounds, respectively.\label{fig:hist3x5}}  
\end{figure}  
\begin{figure}[t]  
 \epsfig{figure=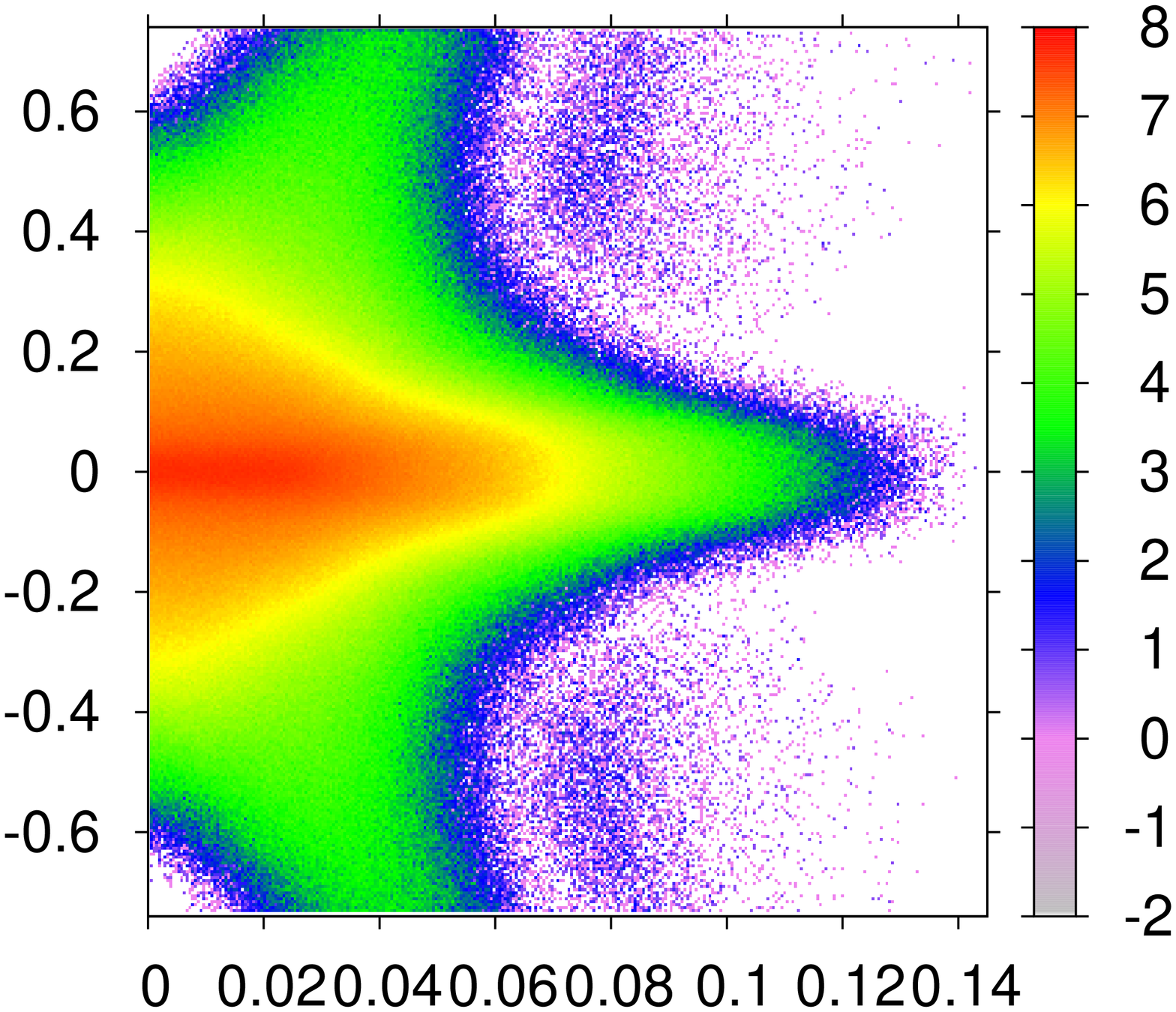,angle=270,width=0.49\textwidth}  
\hfill  
\epsfig{figure=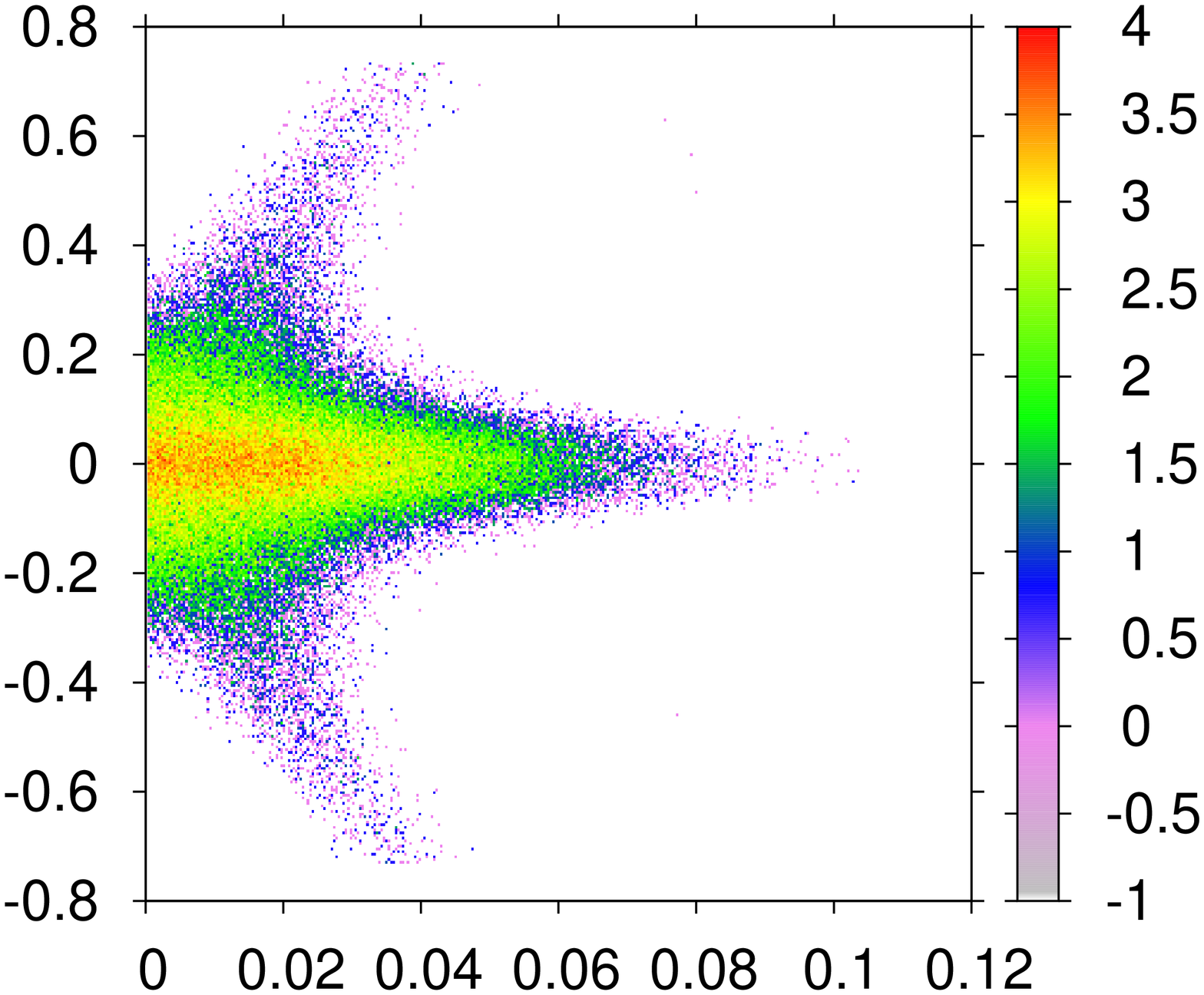,angle=270,width=0.49\textwidth}  
 \epsfig{figure=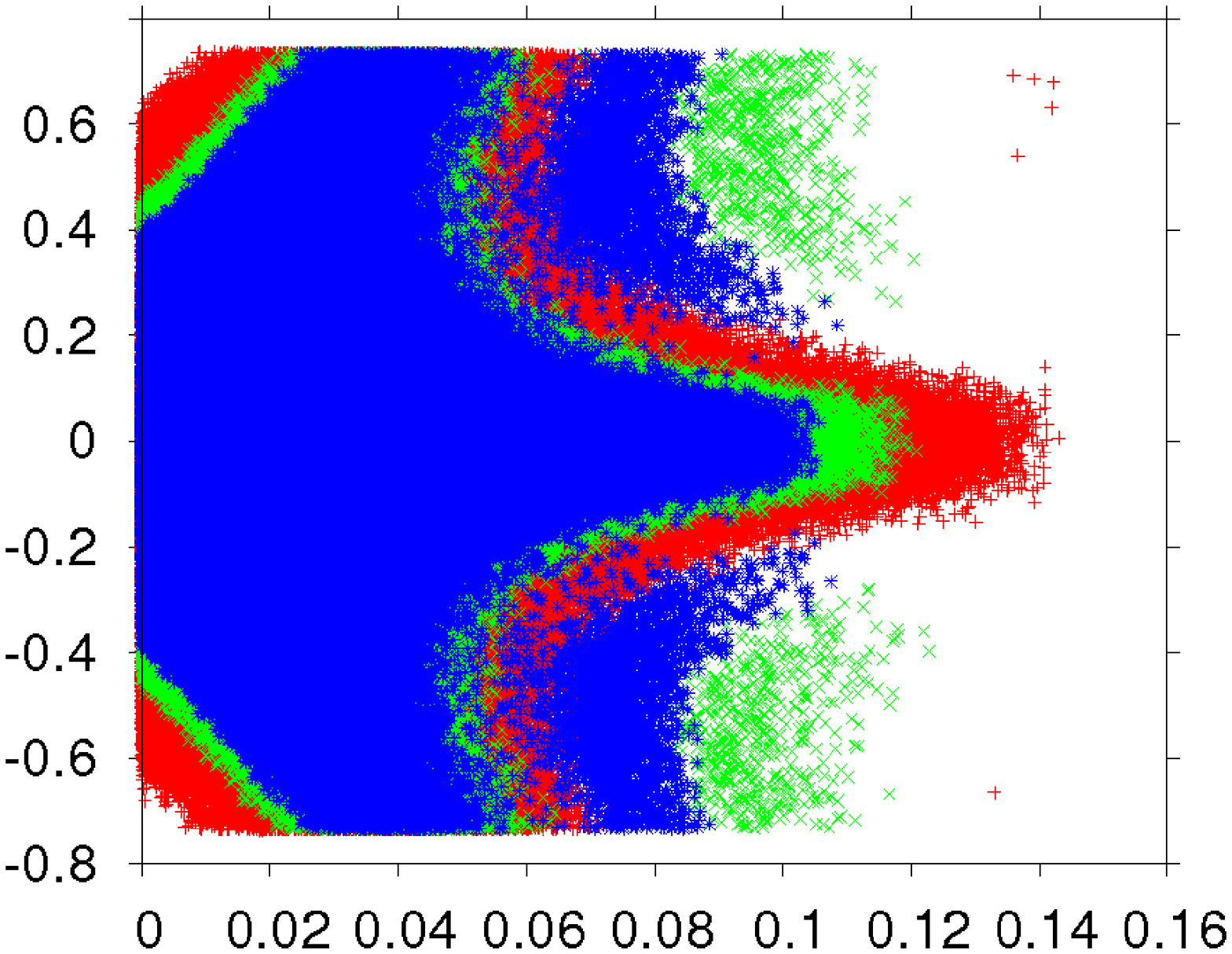,angle=270,width=0.49\textwidth}  
\hfill  
\epsfig{figure=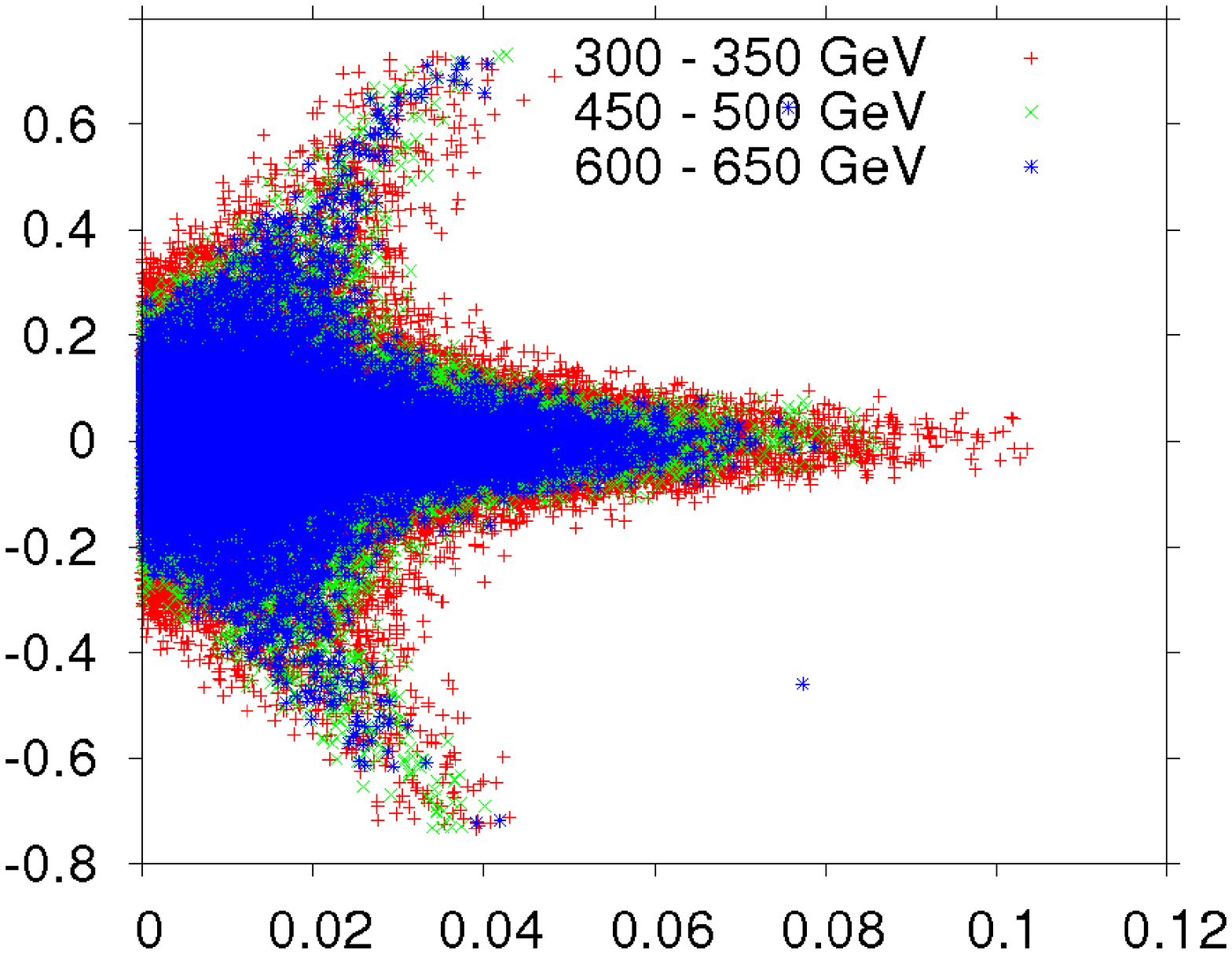,angle=270,width=0.49\textwidth}  
\caption{The allowed parameter ranges  in $\theta_{24}$ and $\theta_{34}$. For further explanation, see the caption of figure \ref{fig:hist3x5}.\label{fig:hist5x6}}  
\end{figure}  
\begin{figure}[t]  
 \epsfig{figure=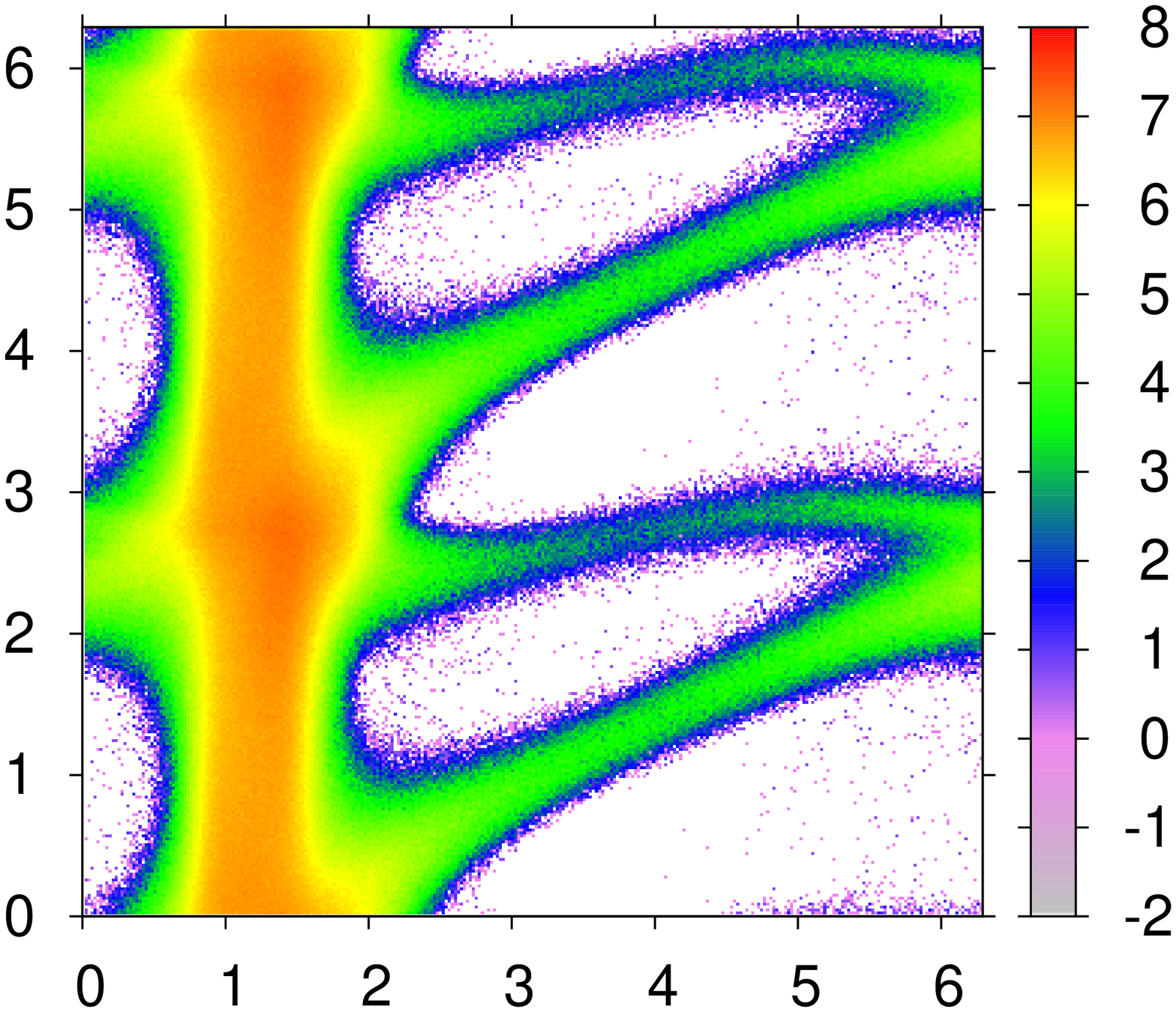,angle=270,width=0.49\textwidth}  
\hfill  
\epsfig{figure=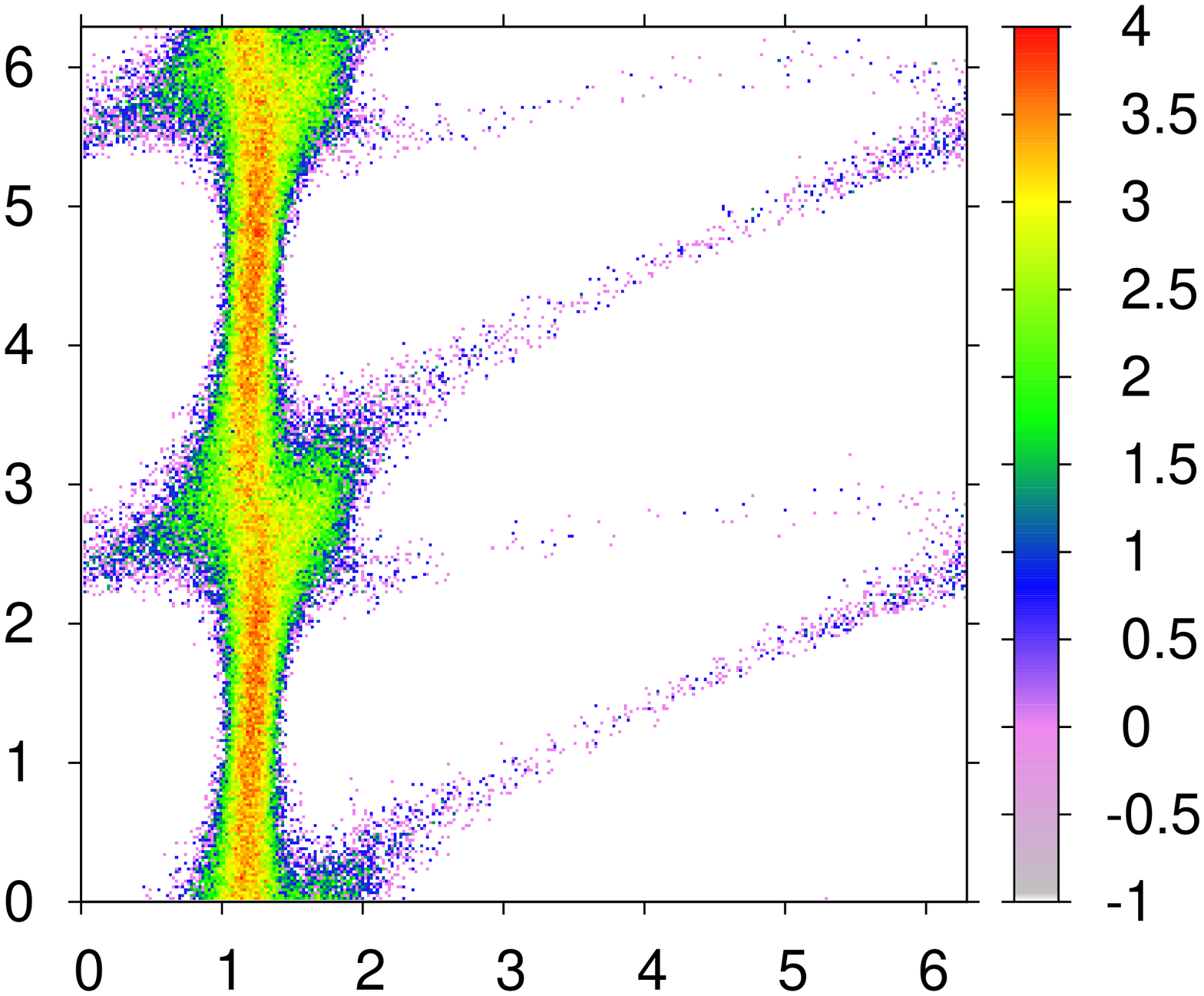,angle=270,width=0.49\textwidth}  
 \epsfig{figure=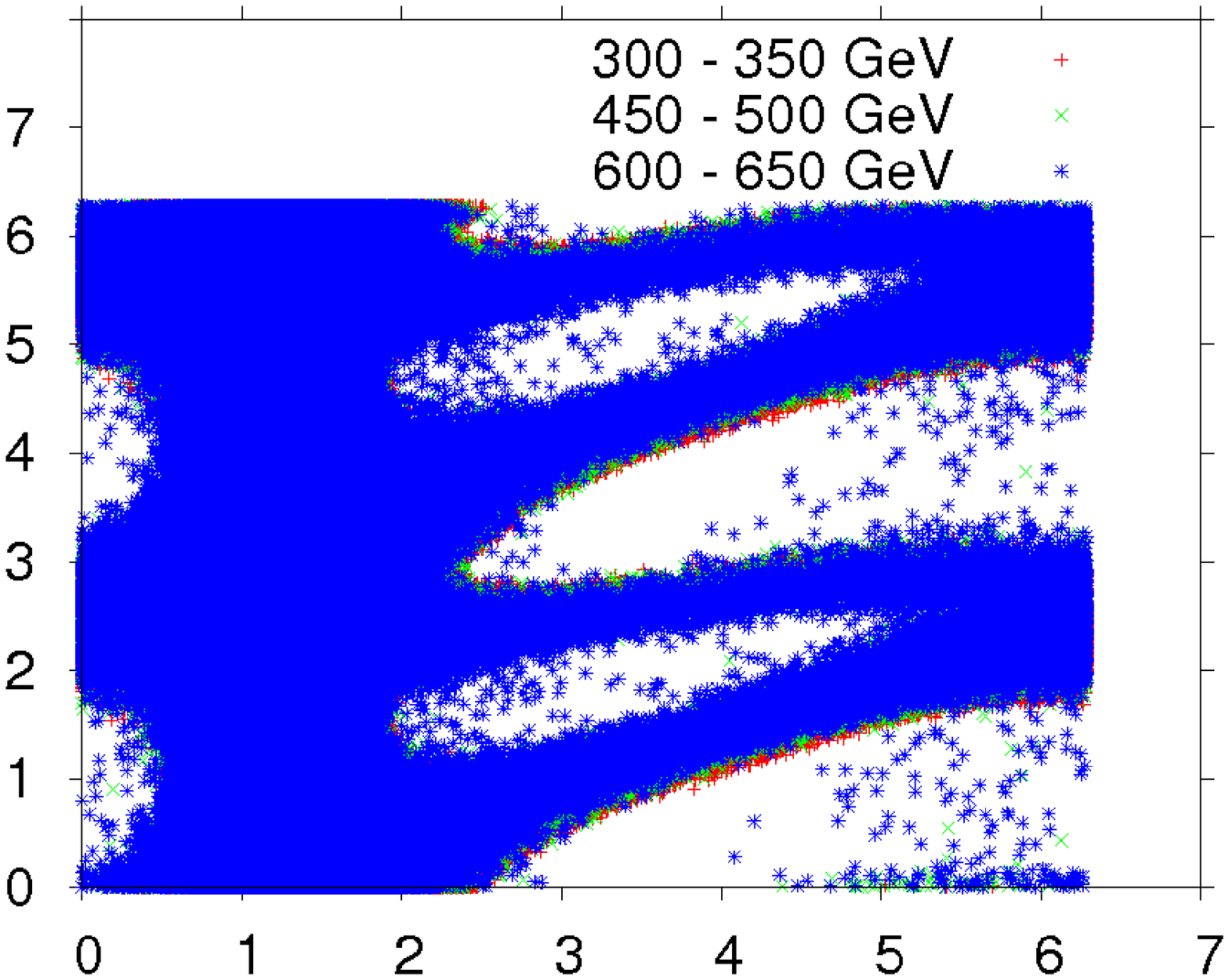,angle=270,width=0.49\textwidth}  
\hfill  
\epsfig{figure=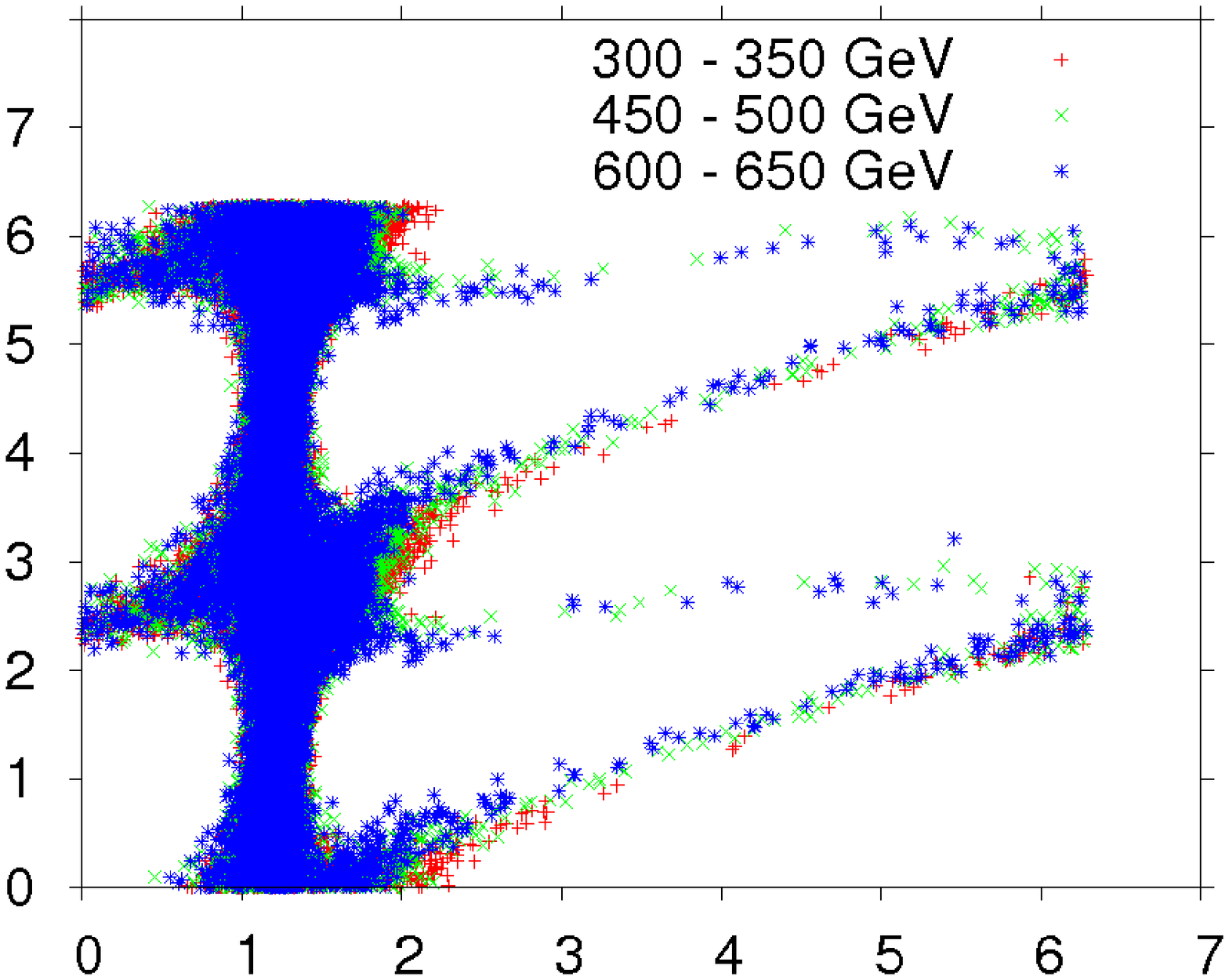,angle=270,width=0.49\textwidth} 
\caption{The allowed parameter ranges  in $\delta_{13}$ and $\delta_{14}$. For further explanation, see the caption of figure \ref{fig:hist3x5}.\label{fig:hist7x8}}  
\end{figure}  
\begin{figure}[t]  
 \epsfig{figure=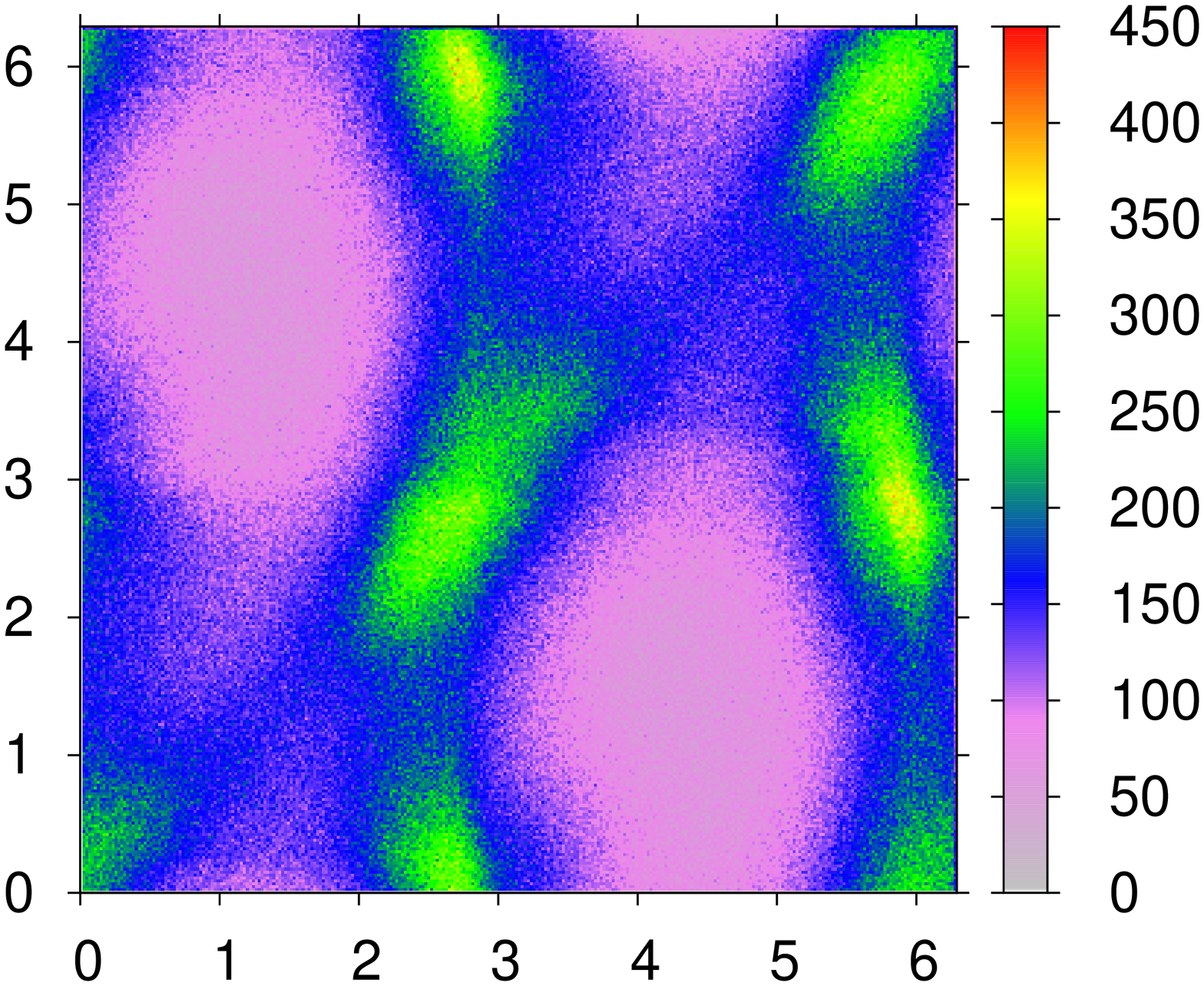,angle=270,width=0.49\textwidth}  
\hfill  
\epsfig{figure=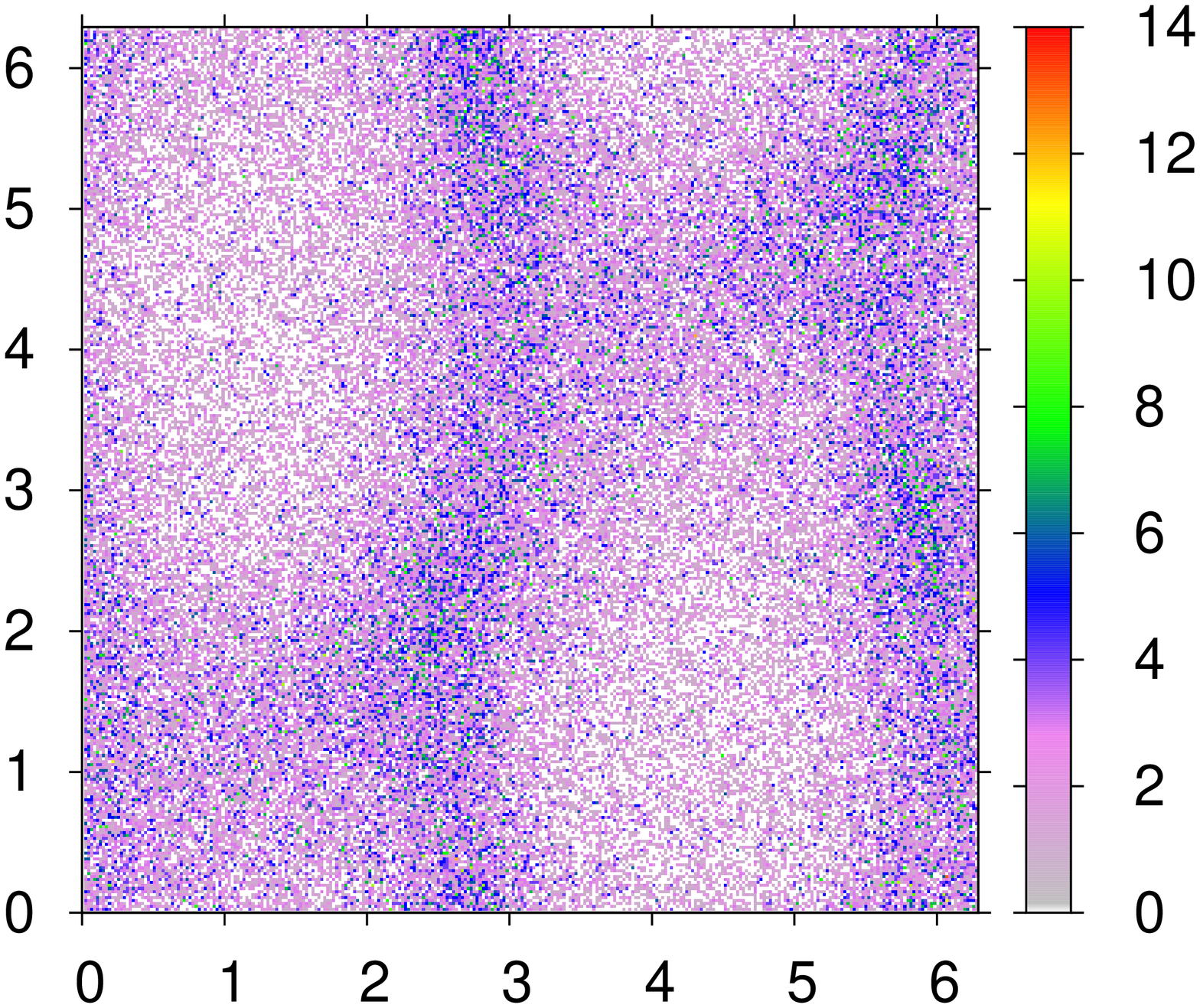,angle=270,width=0.49\textwidth}  
\caption{The allowed parameter ranges  in $\delta_{14}$ and $\delta_{24}$. For further explanation, see the caption of figure \ref{fig:hist3x5}. Here, the mass dependence is not explicitly shown, as all combinations are allowed for each mass $m_{t'}$.\label{fig:hist8x9}}  
\end{figure}  
\begin{figure}[t] 
 \epsfig{figure=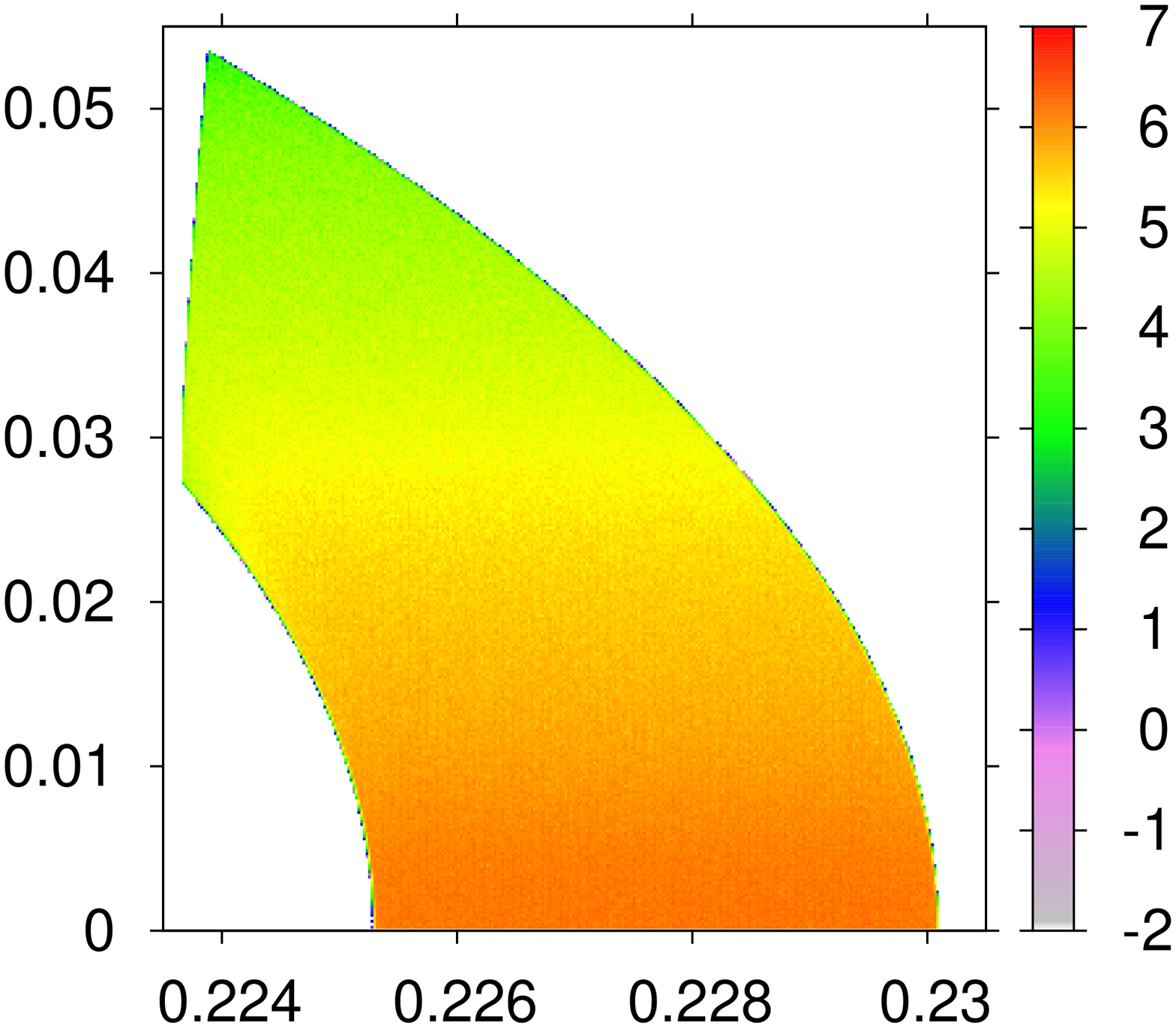,angle=270,width=0.49\textwidth} 
\hfill 
\epsfig{figure=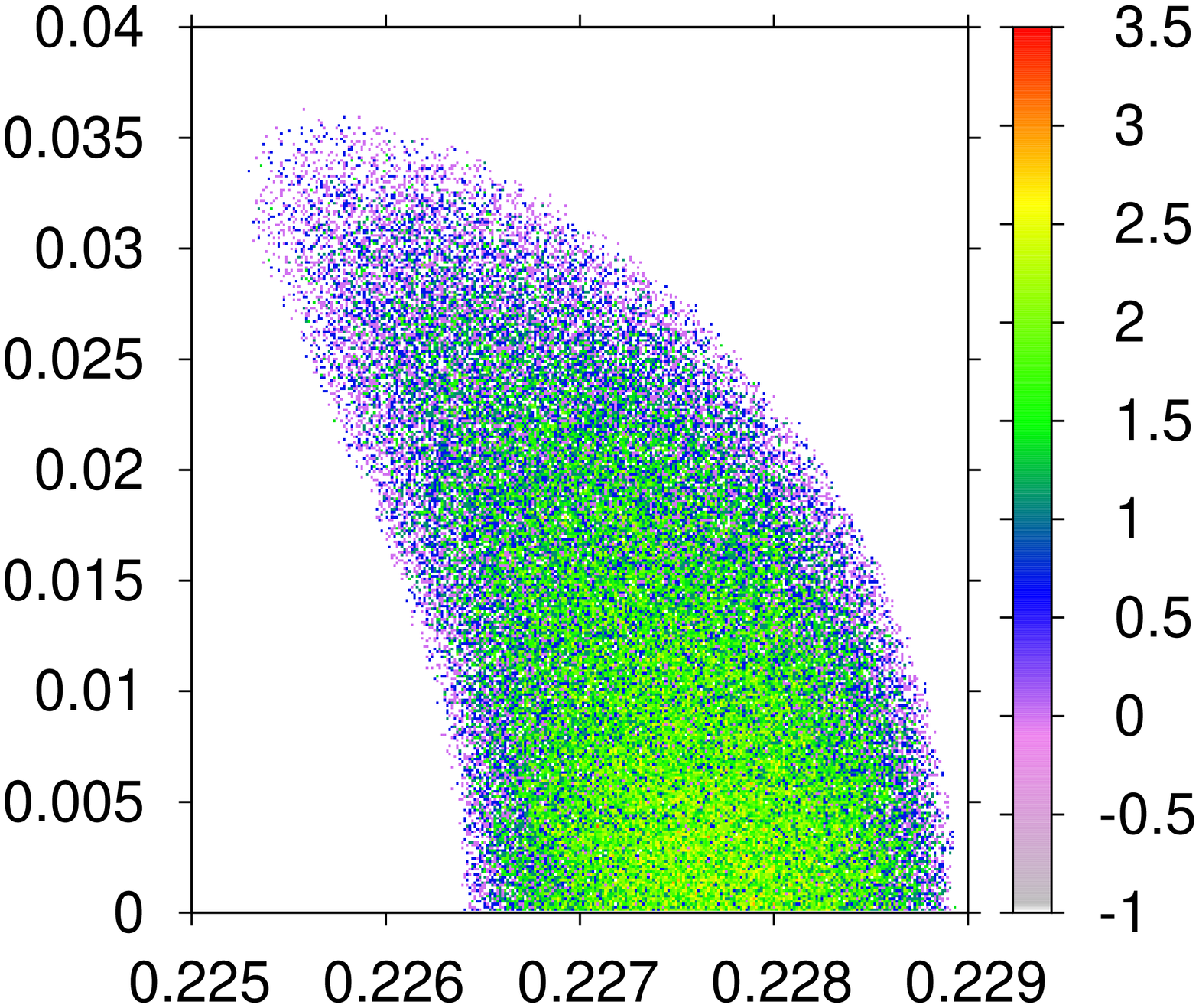,angle=270,width=0.49\textwidth} 
\caption{The allowed parameter ranges  in $\theta_{12}$ and $\theta_{14}$. For further explanation, see the caption of figure \ref{fig:hist3x5}. Here, the mass dependence is not explicitly shown, as all combinations are allowed for each mass $m_{t'}$.\label{fig:hist1x3}} 
\end{figure} 
\begin{figure}[t]  
 \epsfig{figure=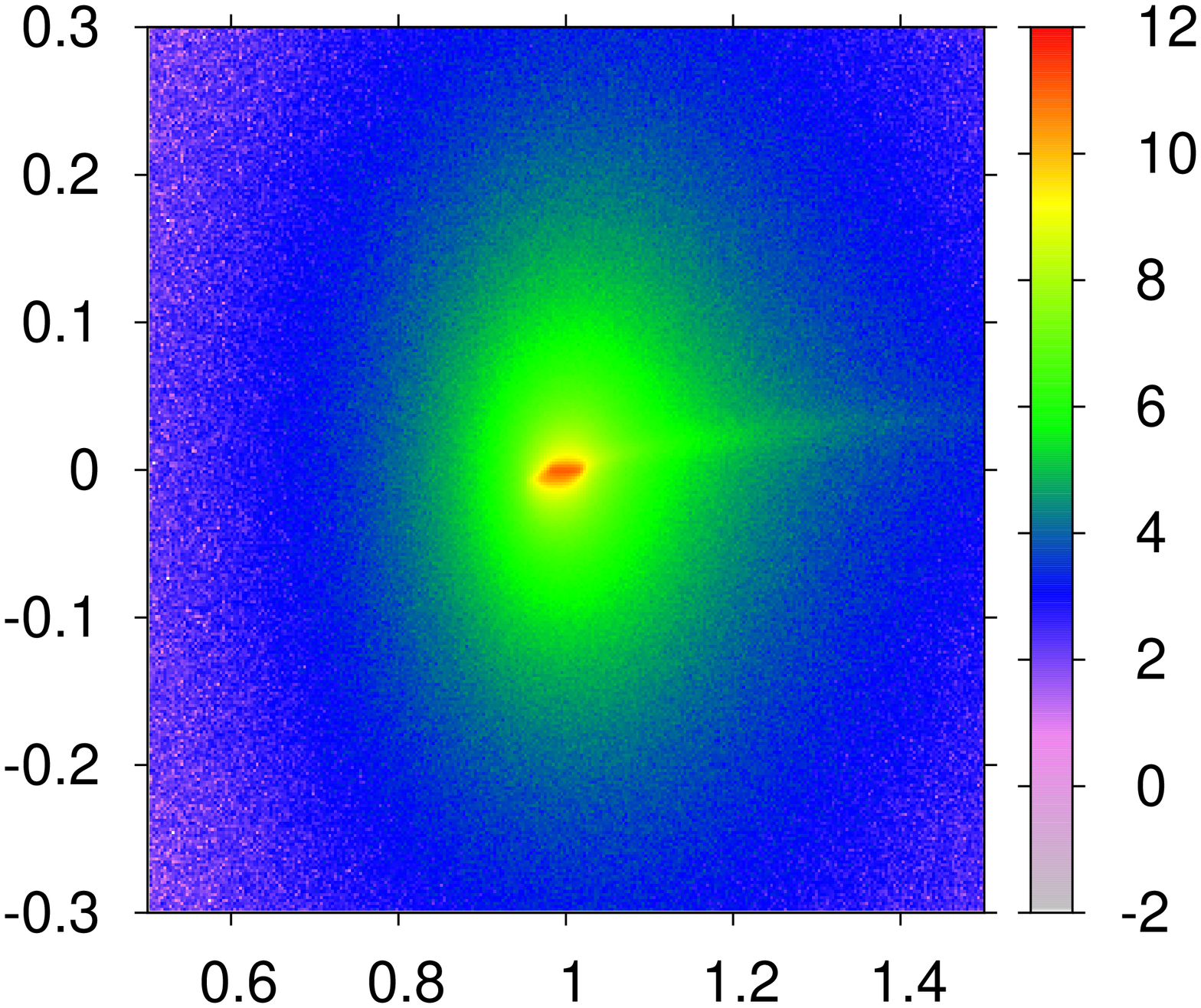,angle=270,width=0.49\textwidth}  
\hfill
\epsfig{figure=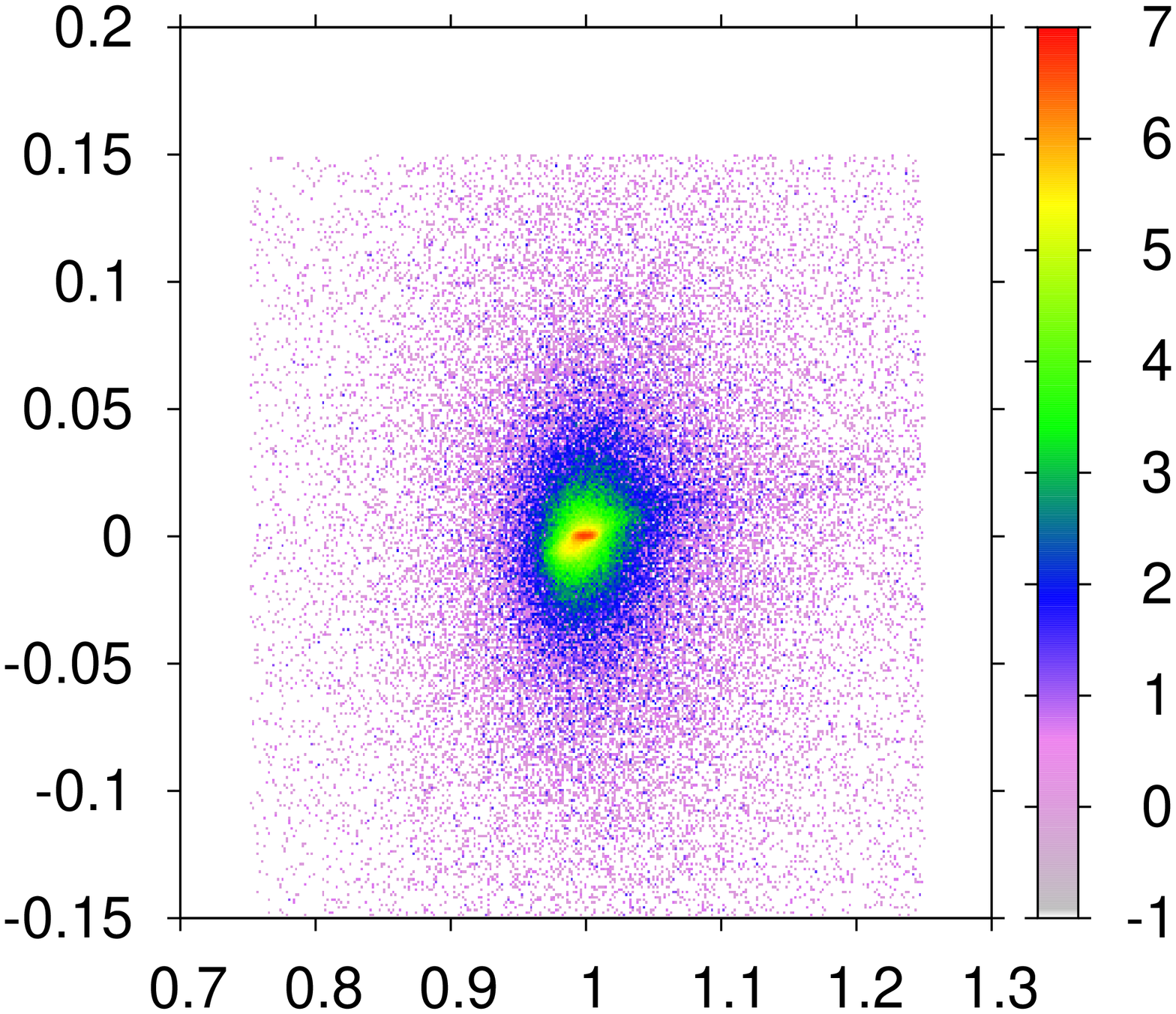,angle=270,width=0.49\textwidth}  
\caption{The results for $\Delta_{K^0}$ shown in the complex plane (real part on the x axis and imaginary part on the y axis): in the left panel for the conservative bounds and in the right panel for the aggressive ones. \label{fig:K0}}  
\end{figure}  
  
\begin{figure}[t]  
 \epsfig{figure=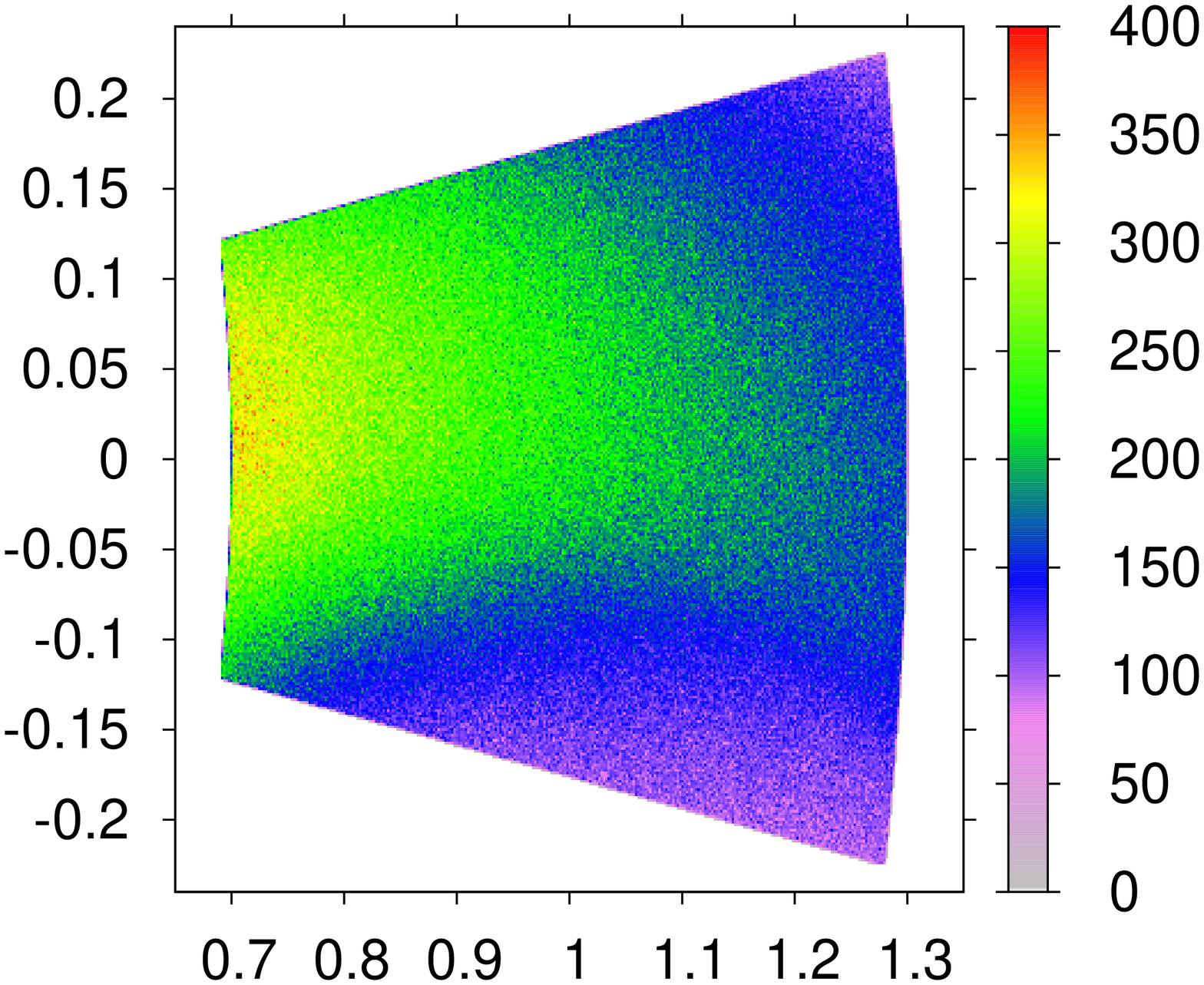,angle=270,width=0.49\textwidth}  
\hfill
\epsfig{figure=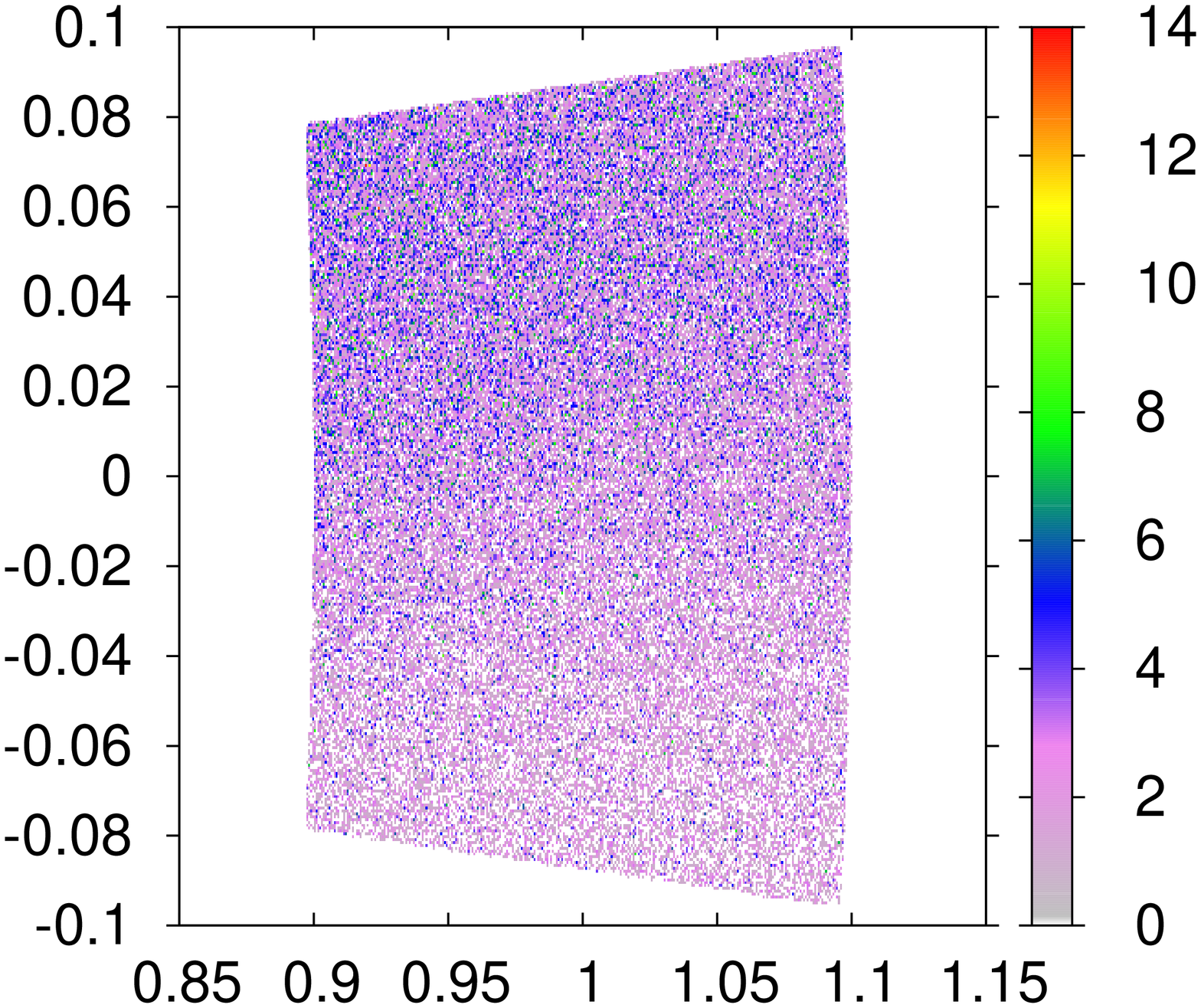,angle=270,width=0.49\textwidth}  
\caption{The results for $\Delta_{B_d}$ as described in the caption of figure \ref{fig:K0}.\label{fig:Bd}  }  
\end{figure}  
  
\begin{figure}[t]  
 \epsfig{figure=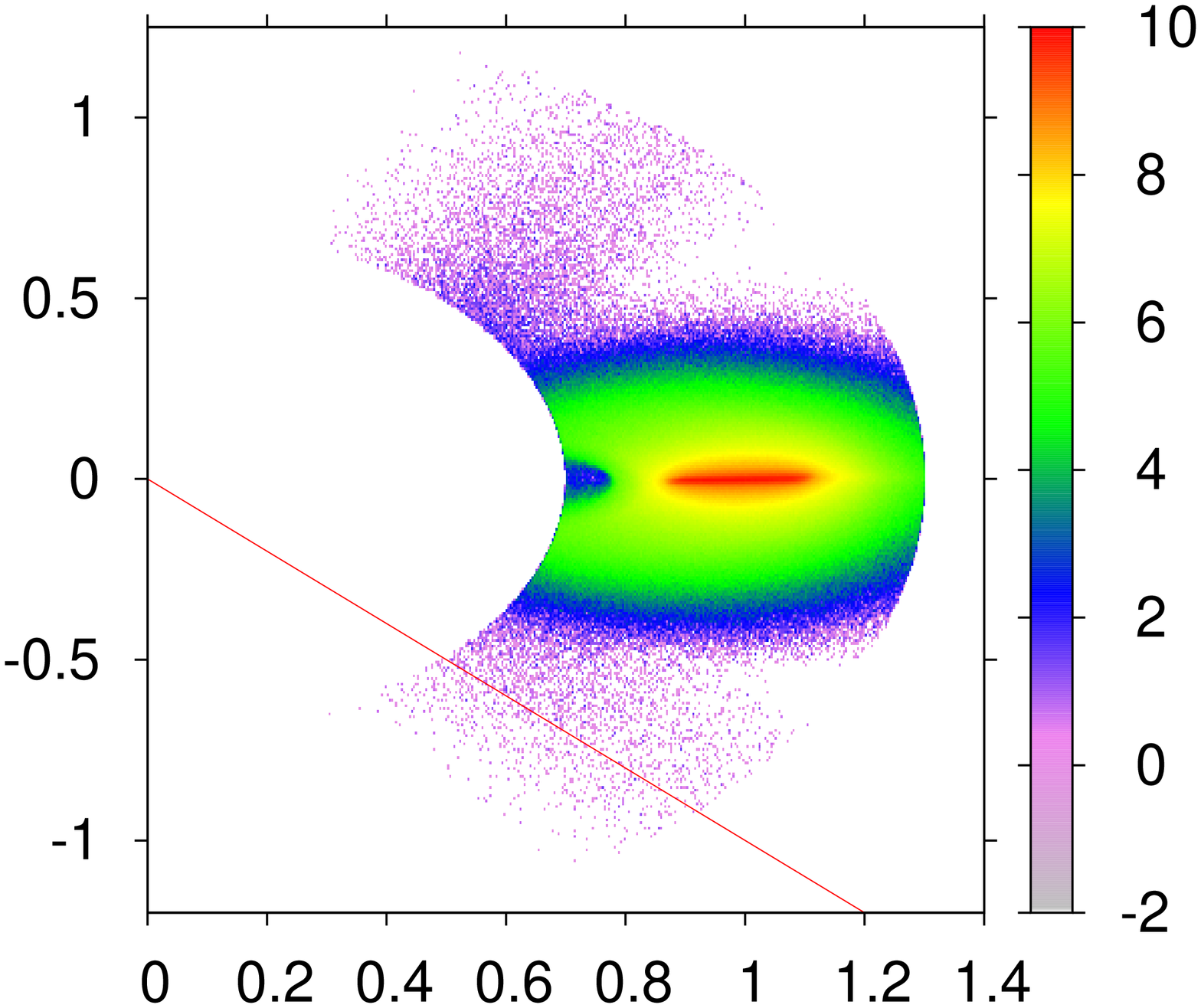,angle=270,width=0.49\textwidth}  
\hfill 
\epsfig{figure=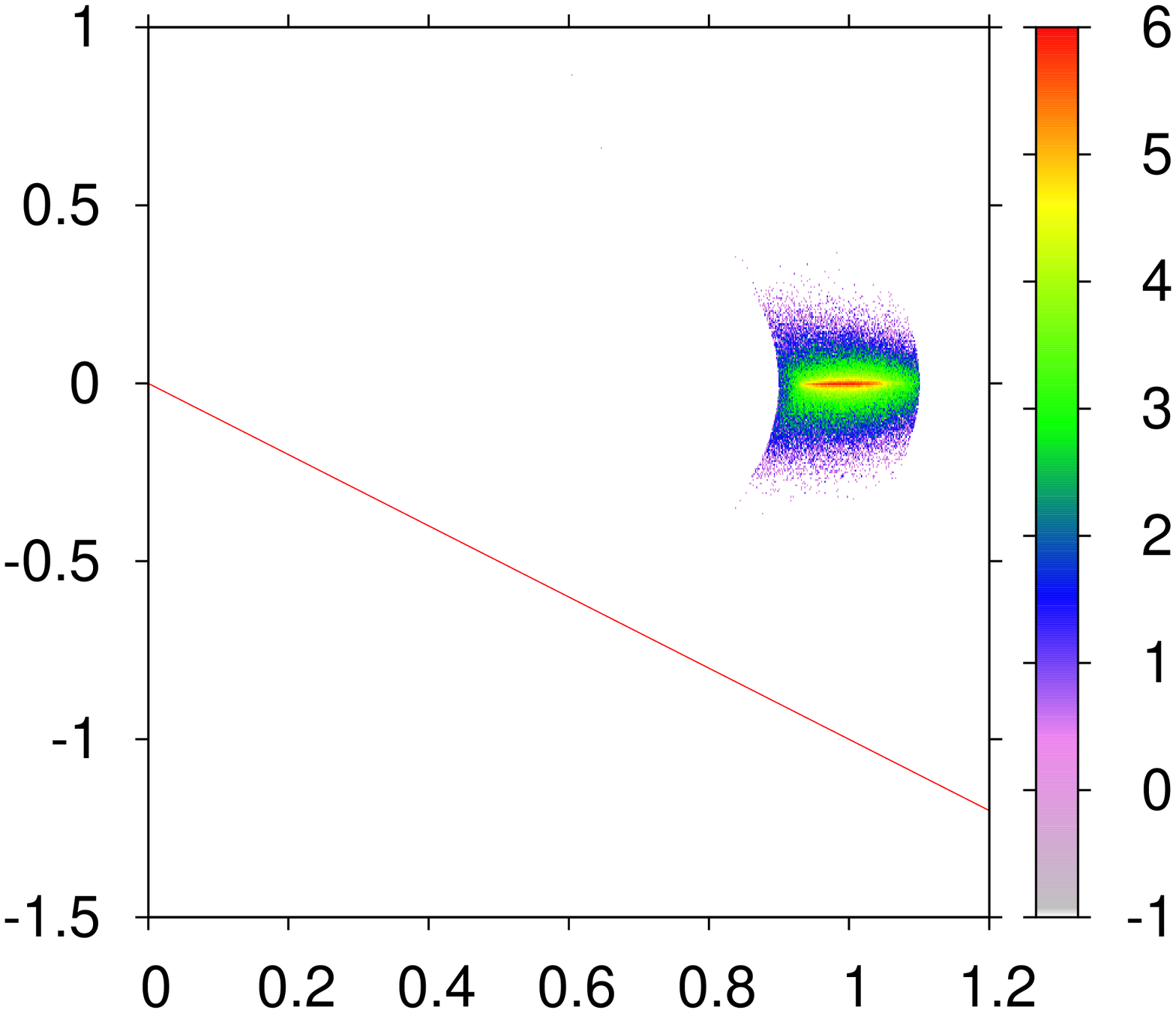,angle=270,width=0.49\textwidth}  
\caption{The results for $\Delta_{B_s}$ as described in the caption of figure \ref{fig:K0}. The red line represents a $\Phi_s$ angle of $-45^\circ$ which is
 hinted by recent experiments.\label{fig:Bs}}  
\end{figure}  
\begin{figure}[t]  
 \epsfig{figure=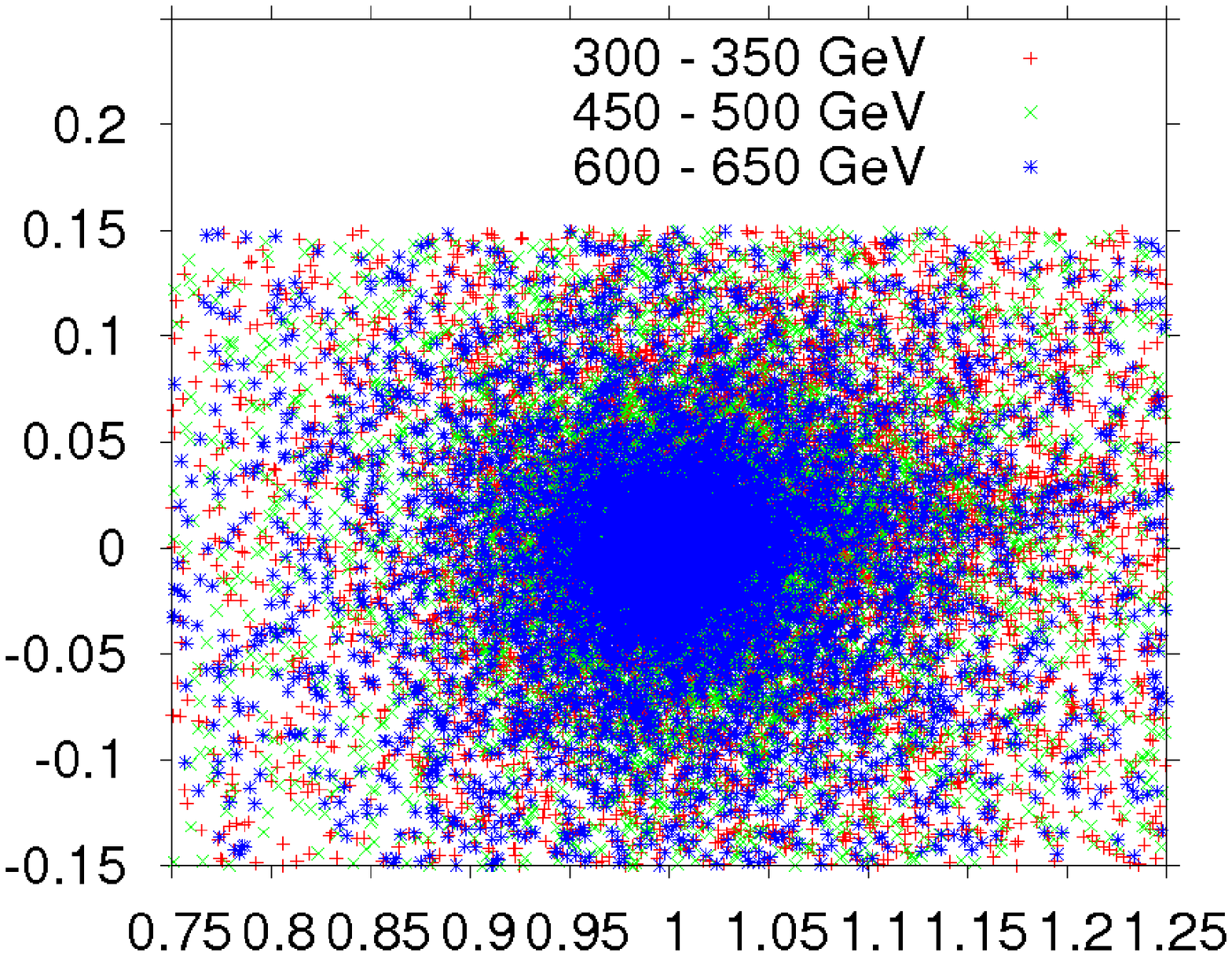,angle=270,width=0.325\textwidth}  
\hfill
\epsfig{figure=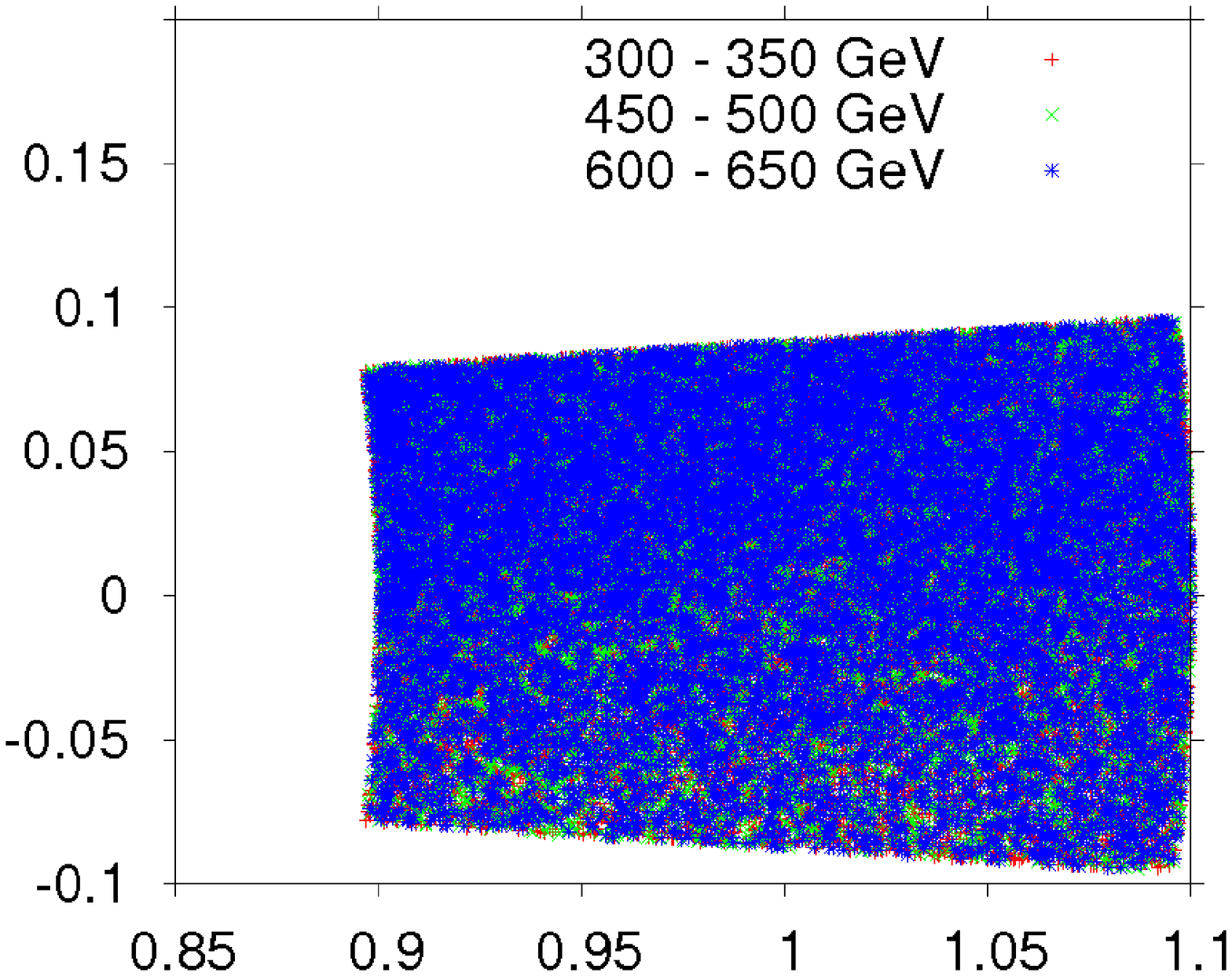,angle=270,width=0.325\textwidth}  
\hfill
\epsfig{figure=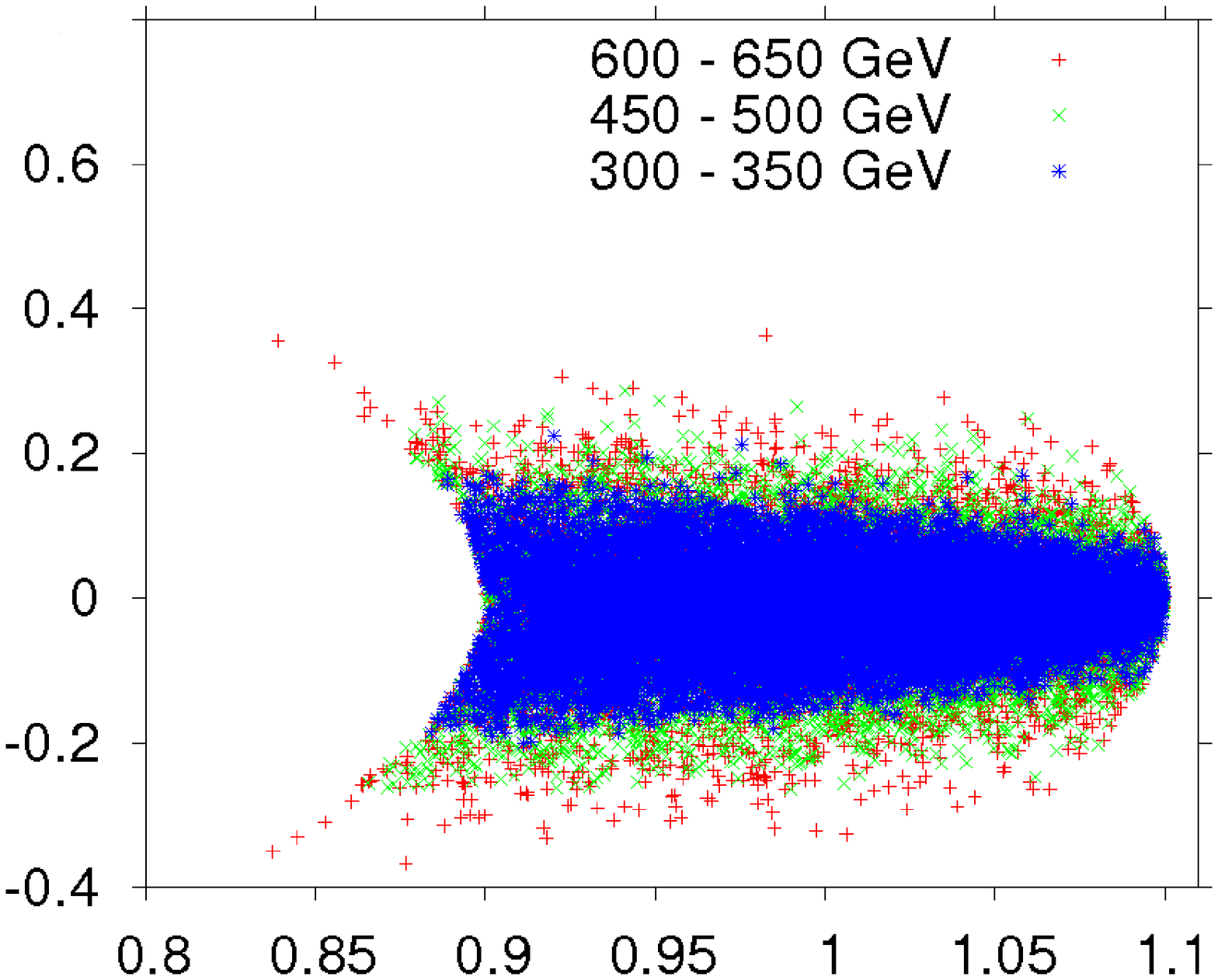,angle=270,width=.325\textwidth}  
\caption{The $m_{t'}$ dependence of the FCNC $\Delta$s: from left to right $\Delta_K$, $\Delta_{B_d}$ and $\Delta_{B_s}$.\label{fig:mdep}}  
\end{figure}  
\begin{figure}[t]  
 \epsfig{figure=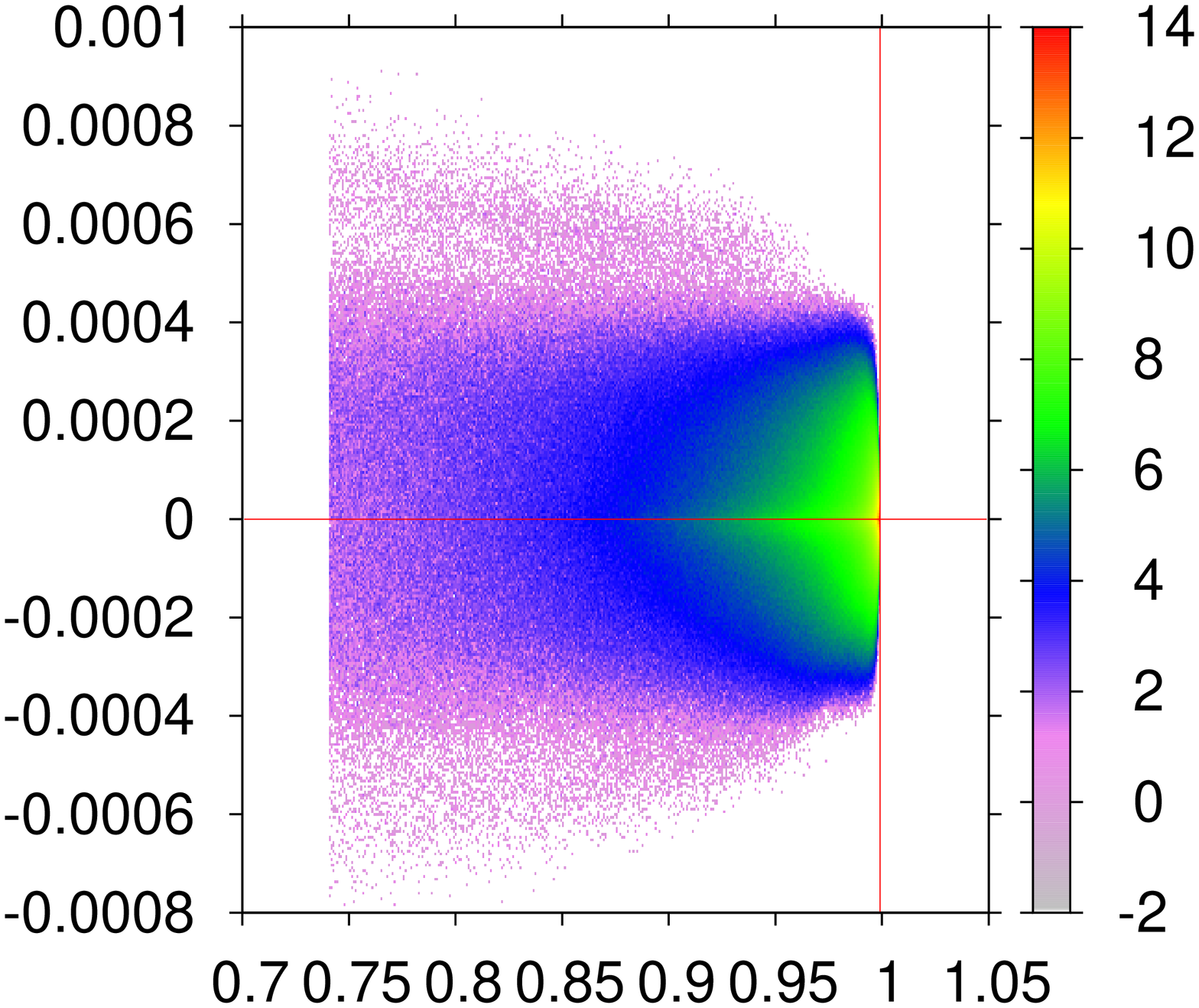,angle=270,width=0.49\textwidth}  
\hfill
\epsfig{figure=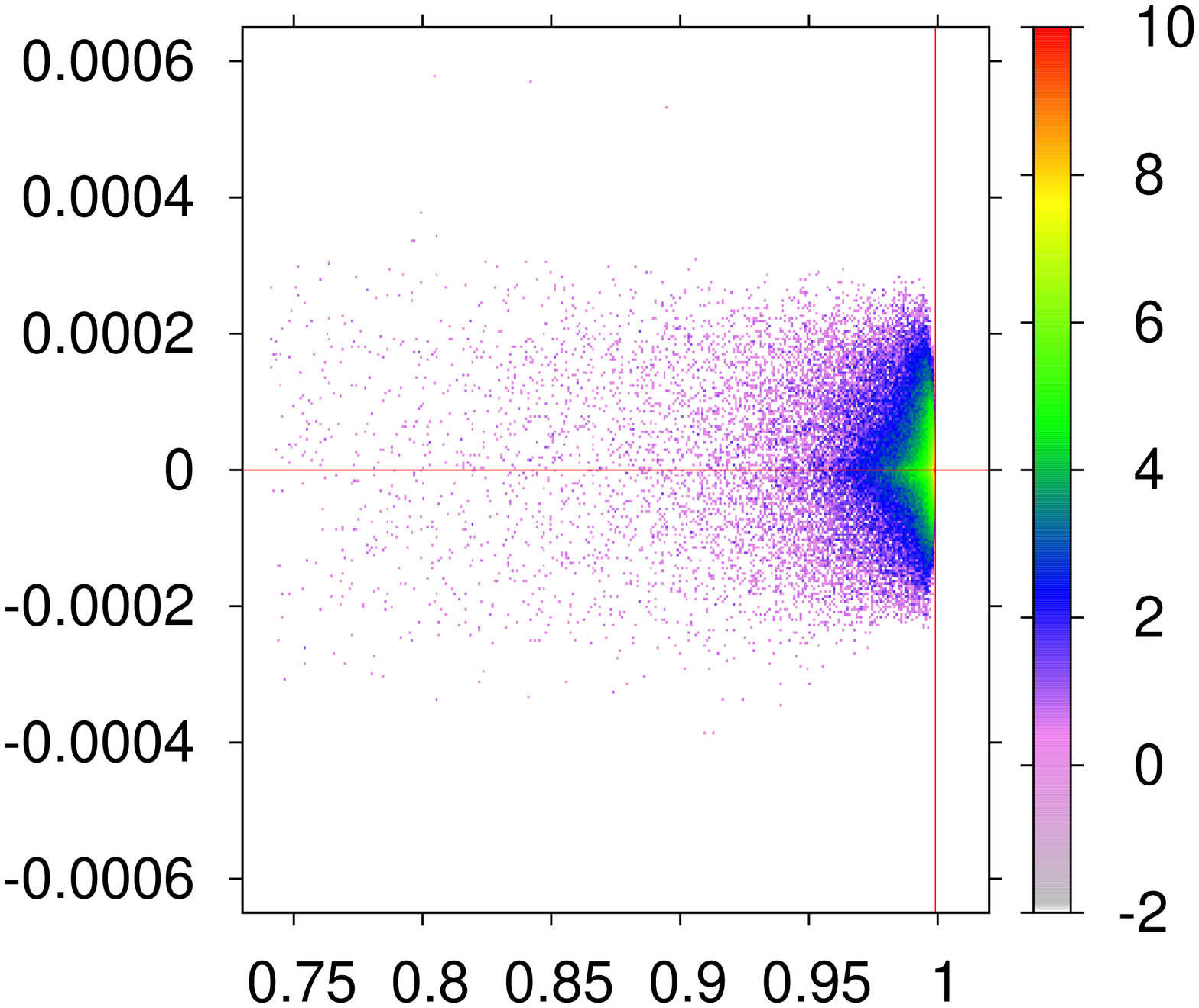,angle=270,width=0.49\textwidth}  
\caption{The results for $V_{tb}$ as described in the caption of figure \ref{fig:K0}. The red crosslines give the value for CKM3.\label{fig:Vtb} }  
\end{figure}  
\begin{figure}[t]  
 \epsfig{figure=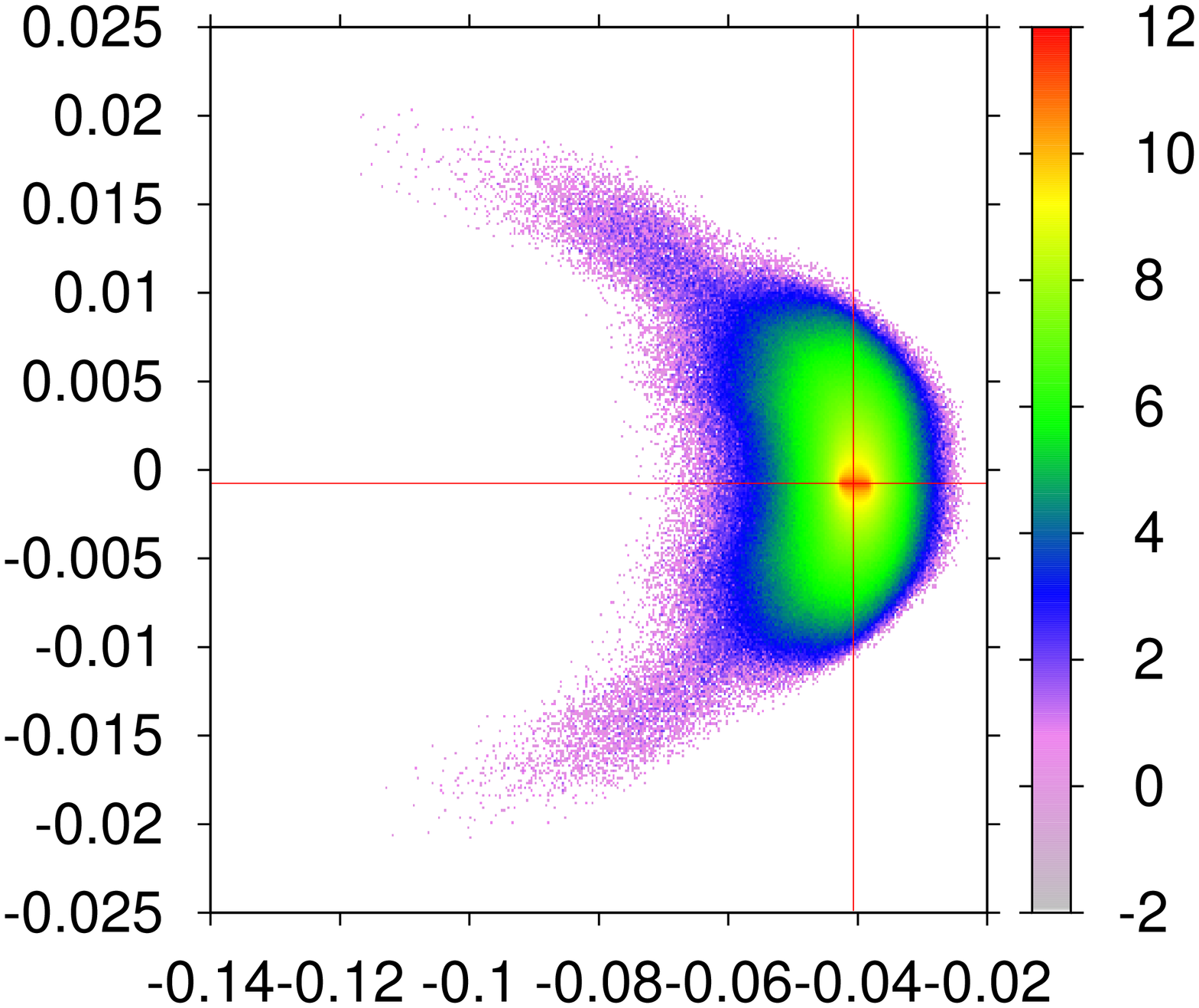,angle=270,width=0.49\textwidth}  
\hfill
\epsfig{figure=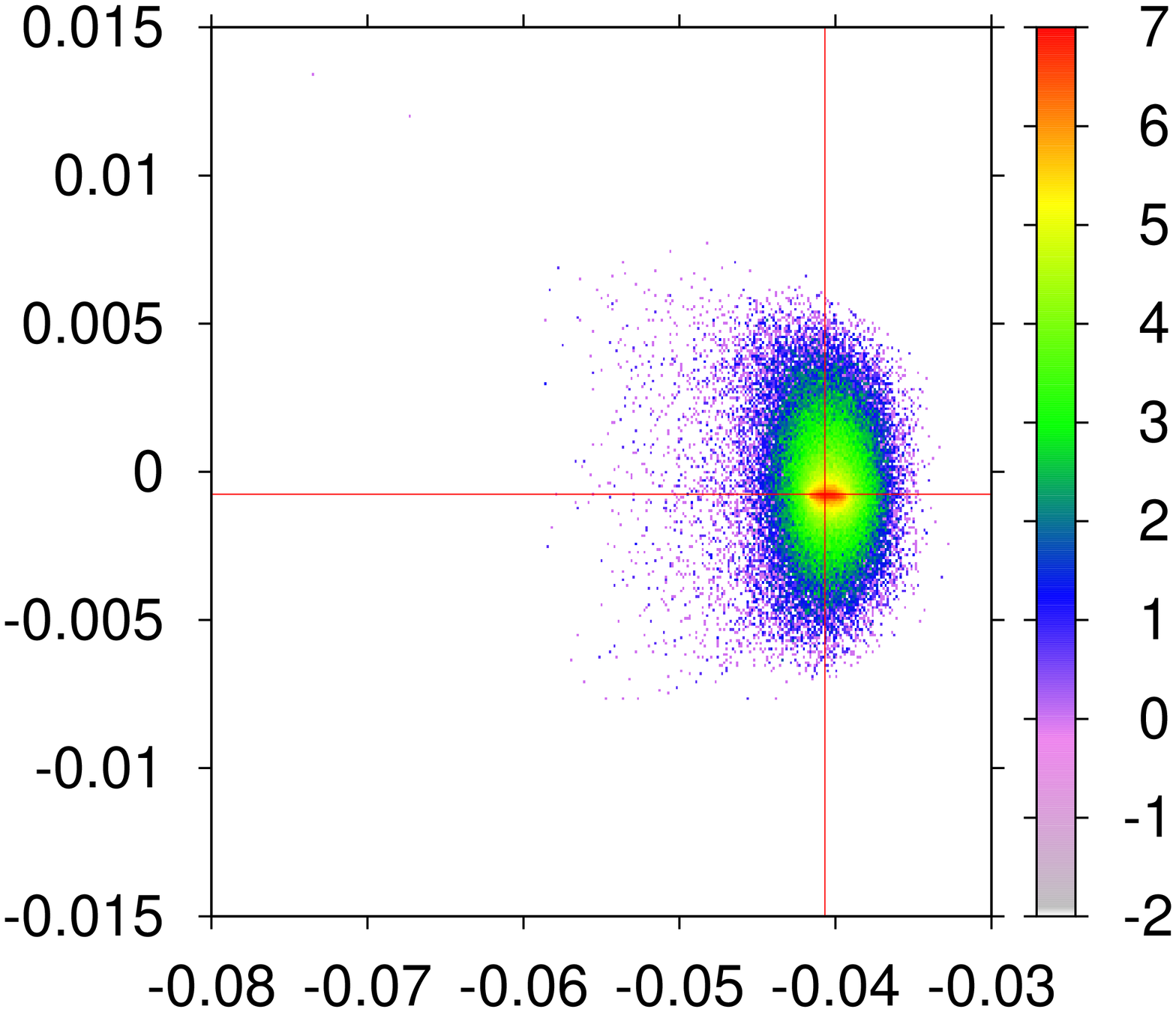,angle=270,width=0.49\textwidth}  
\caption{The results for $V_{ts}$ as described in the caption of figure \ref{fig:K0}.  The red crosslines give the value for CKM3.\label{fig:Vts} }  
\end{figure}  
\begin{figure}[t]  
 \epsfig{figure=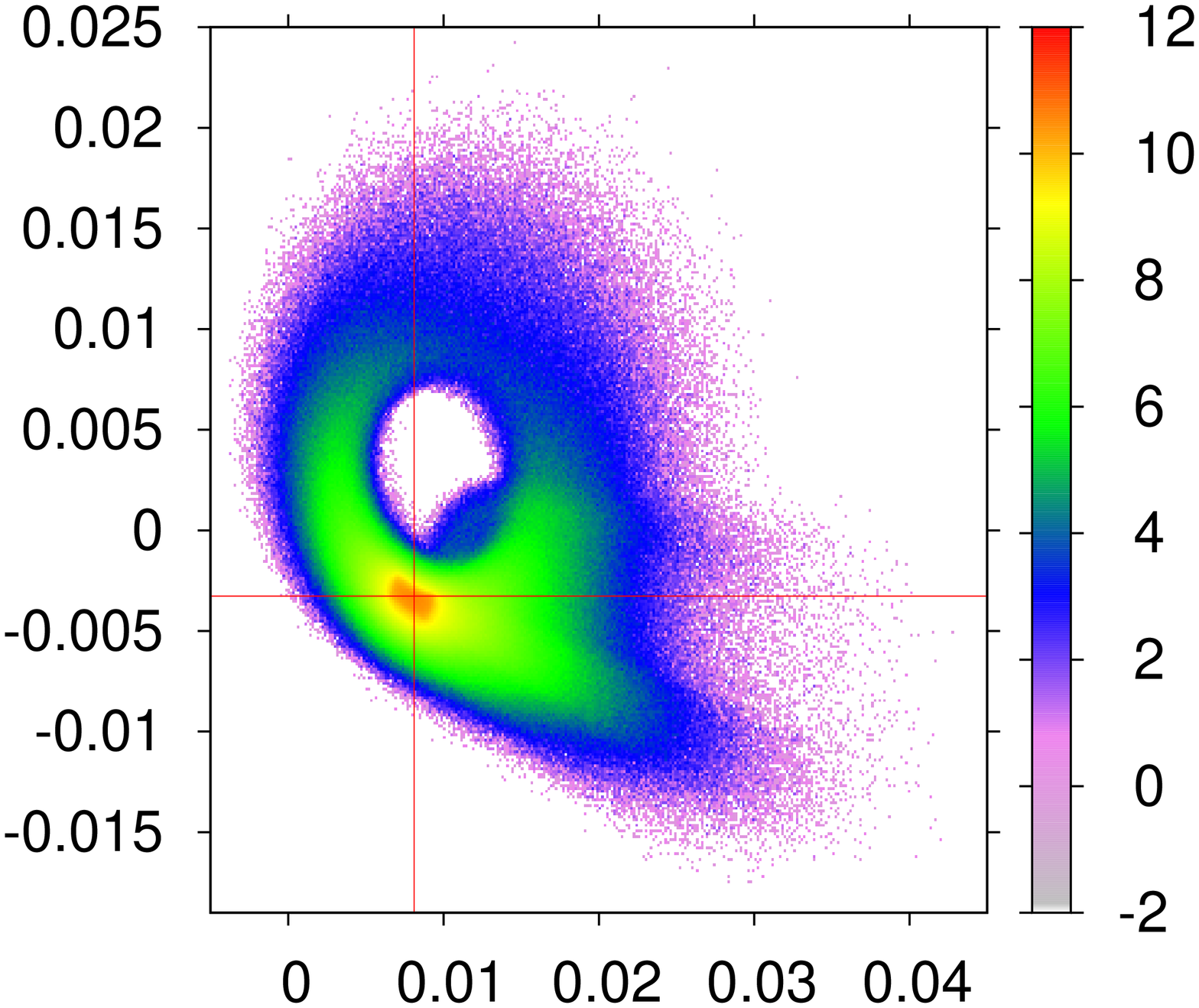,angle=270,width=0.49\textwidth}  
\hfill
\epsfig{figure=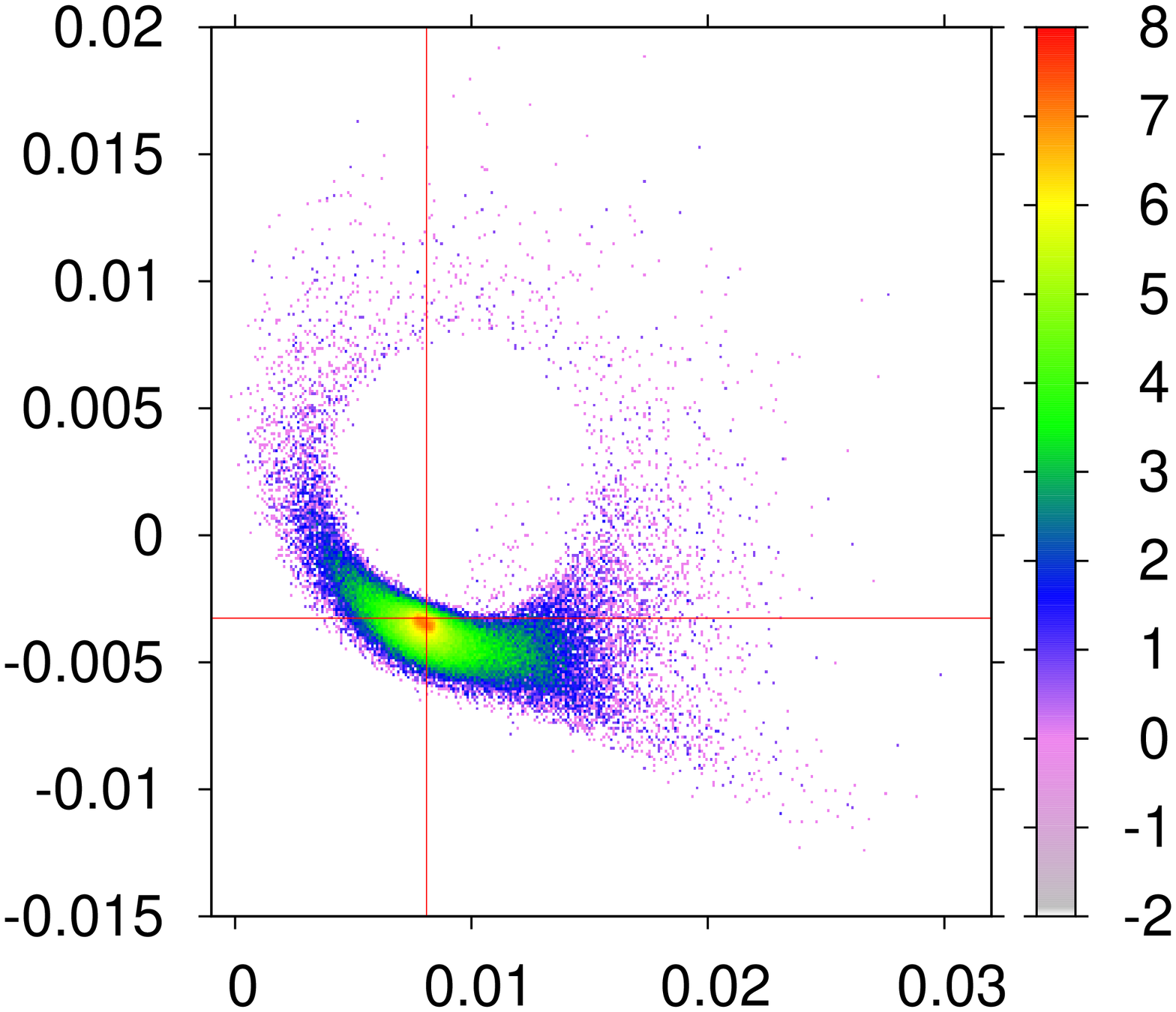,angle=270,width=0.49\textwidth}  
\caption{The results for $V_{td}$ as described in the caption of figure \ref{fig:K0}. The red crosslines give the value for CKM3. The peculiar ring structure already arises after enforcing unitarity and tree-level bounds. \label{fig:Vtd}}  
\end{figure}  
\begin{figure}[t]  
 \epsfig{figure=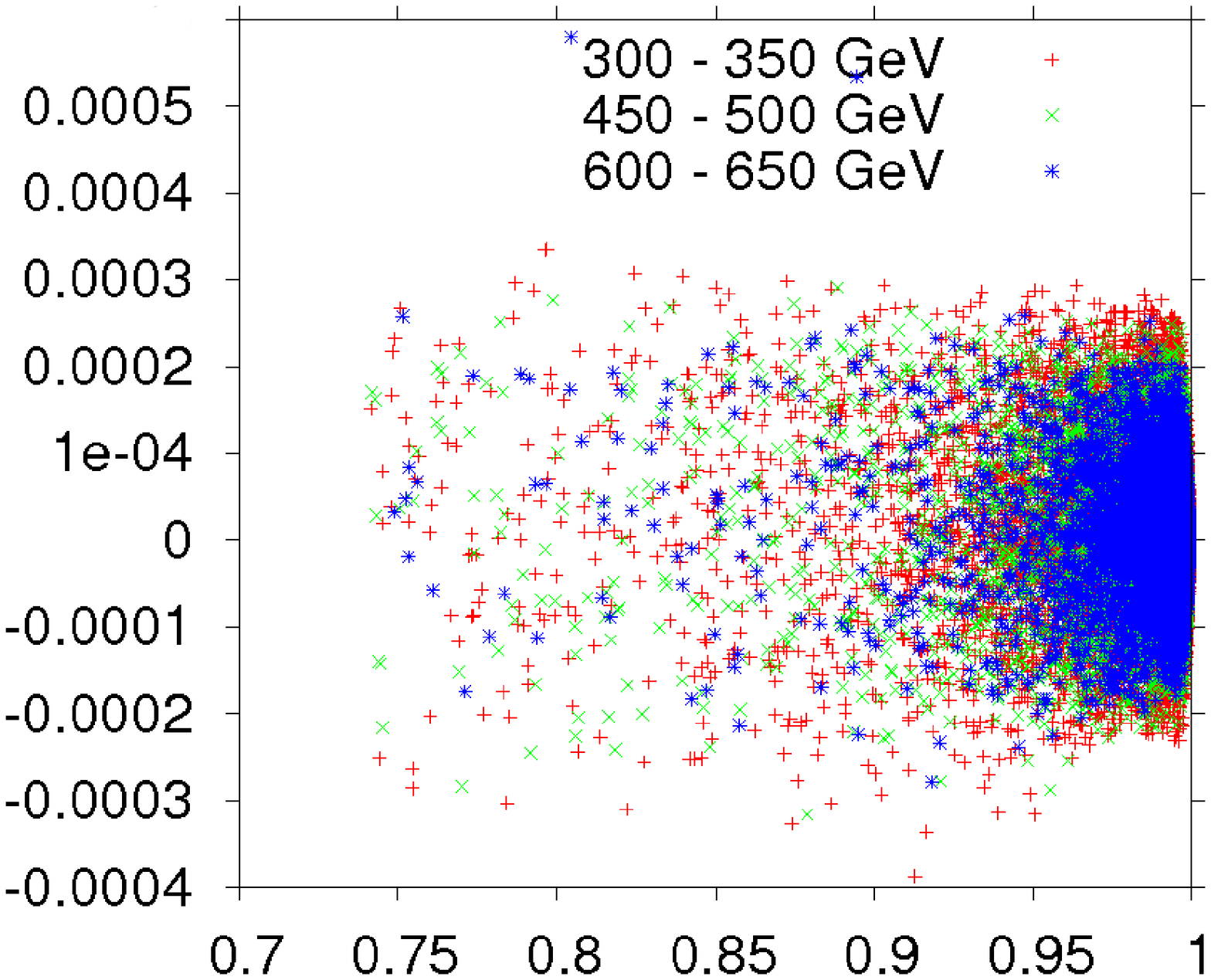,angle=270,width=0.325\textwidth}  
\hfill
\epsfig{figure=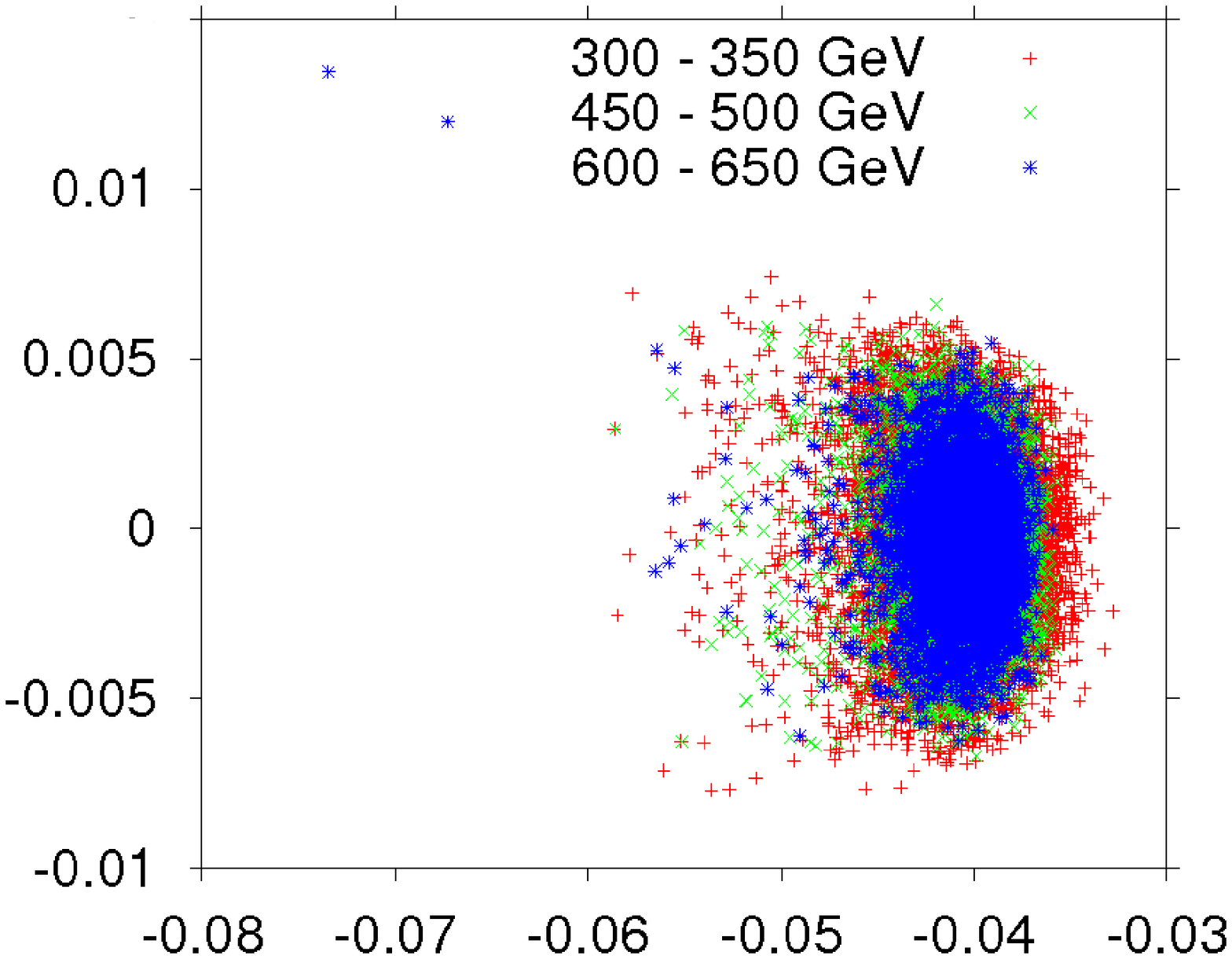,angle=270,width=0.325\textwidth}  
\hfill
\epsfig{figure=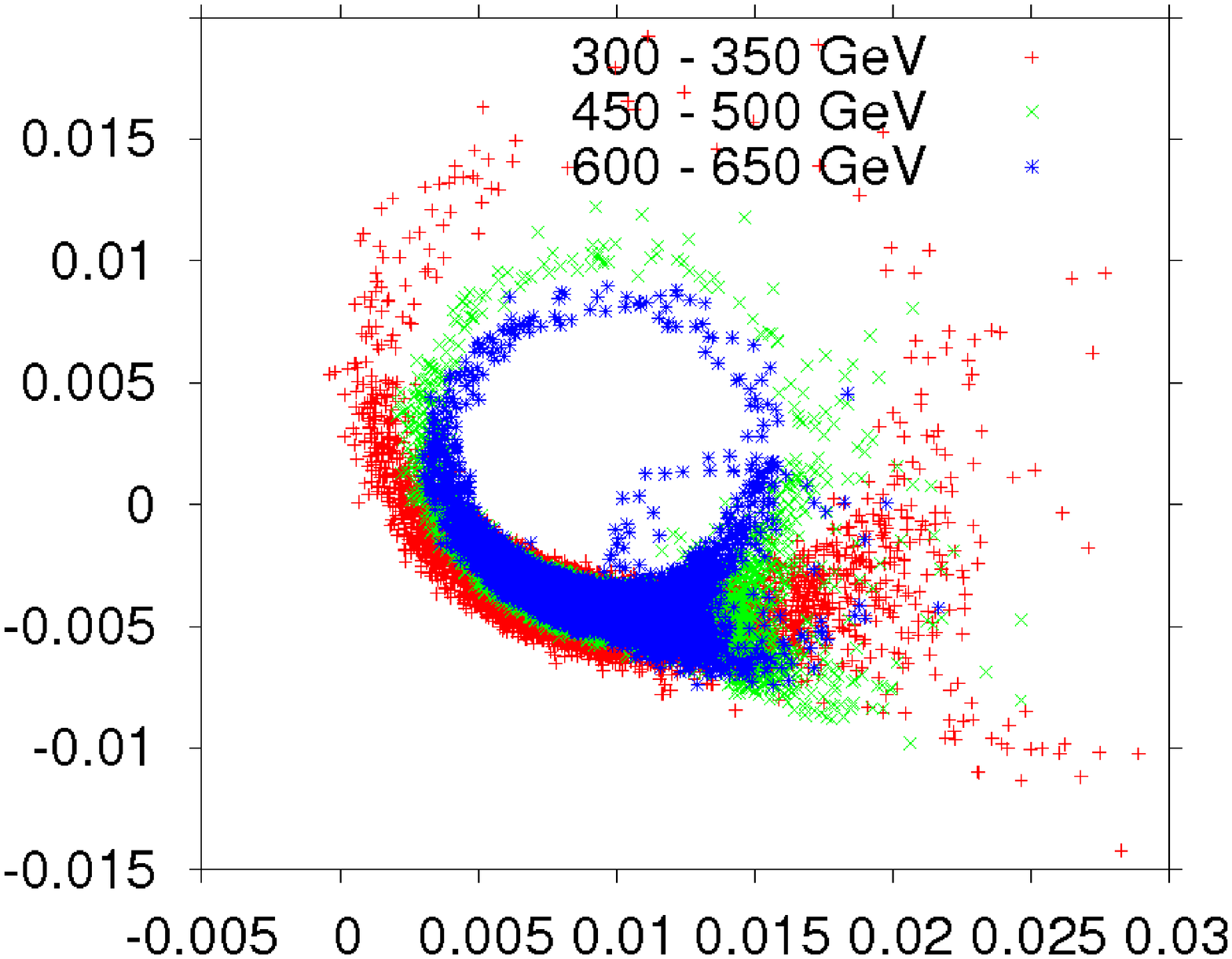,angle=270,width=0.325\textwidth}  
\caption{The dependence of the CKM elements $V_{tb}$ (left panel), $V_{ts}$ (middle panel) and $V_{td}$ (right panel) on the mass region.\label{fig:MassVckm}}  
\end{figure}  
\begin{figure}[t]  
 \epsfig{figure=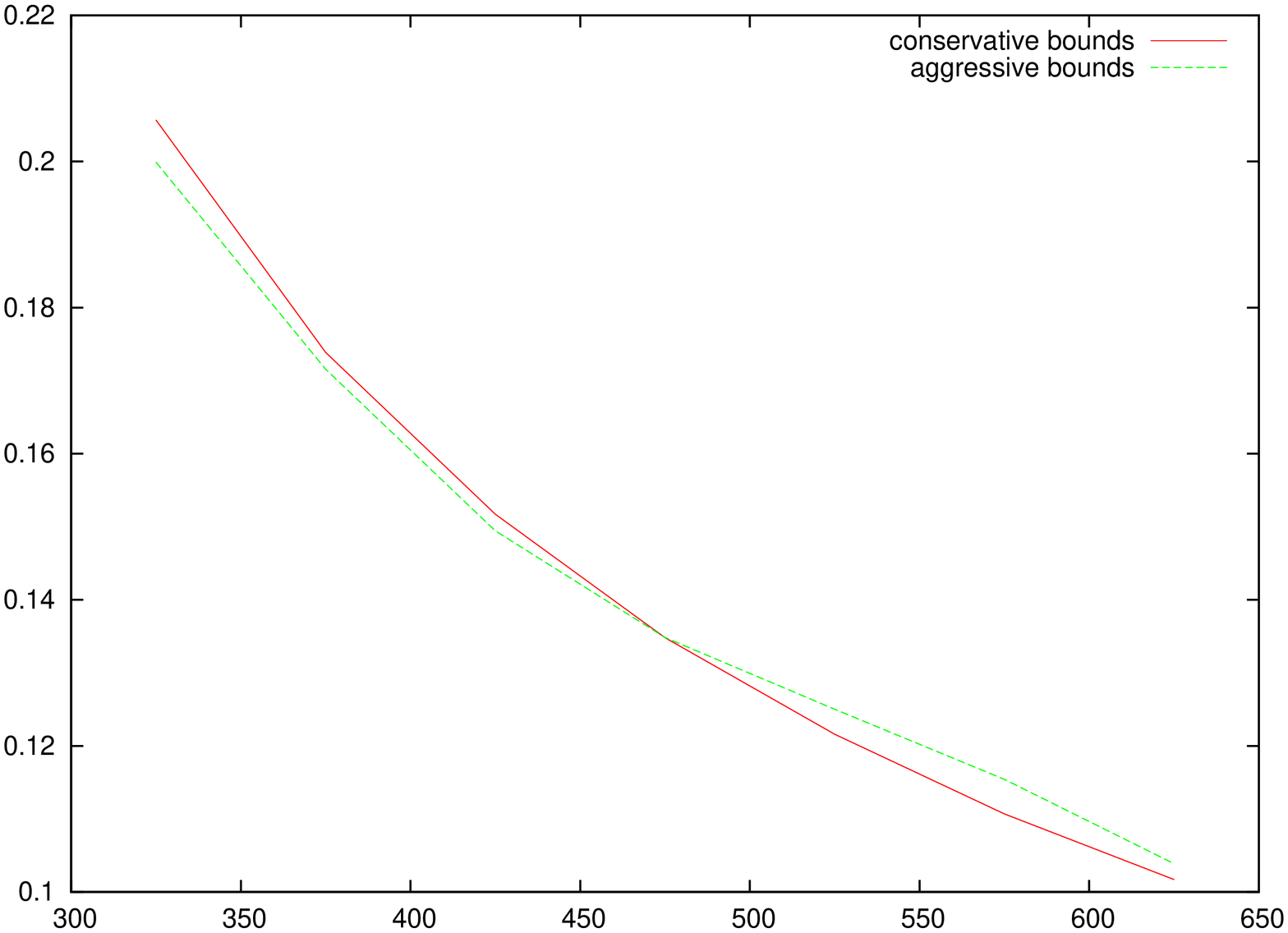,angle=270,width=\textwidth}  
\caption{Relative distribution of the accepted $12\,817\,846$ and $150\,763$ points using  the conservative and the aggressive bounds, respectively. The relative occurrence is shown on the y axis and $m_{t'}$ on the x axis.\label{fig:mdist}}  
\end{figure}  
Subsequently, we will describe the  scan through the nine-dimensional parameter space of the 4$\times$4 mixing matrix   
and the mass regions for $m_{t'}$  and test whether the experimental constraints on quark mixing  
are fulfilled.   For this purpose we use the exact parameterization of $V_{CKM4}$ described in  
Section 2.1, Eq. \eqref{eq:CKM4FP}. 
For the allowed ranges -- especially  on the new parameters related to the fourth generation --  
it is crucial how to treat the errors of  the tree level bounds.   
We have decided to study two different treatments of the error ranges.  We adopt a   
conservative and an aggressive set of bounds. In both the conservative and the aggressive   
case, the bound on $V_{tb}$ is assumed to be hard. We enforce each of  the
six other tree-level constraints  
to be individually fulfilled at the 2$\sigma$ level,  
 i.e. our CKM matrix element $V_{\text{CKM4},i}$ has to be in the range   
$$|V_i|-2\Delta V_i<|V_{\text{CKM4},i}|< |V_i|+2\Delta V_i.$$ Additionally, in order to have   
a measure for the deviation from the central values of the tree level bounds, we define   
a $\chi^2$ per degree of freedom (d.o.f.) as   
$$\chi^2/{\text{d.o.f.}}=\frac{1}{n}\sum_{i=ud,us,ub,cd,cd,cb}\left(\frac{|V_{\text{CKM4},i}|-|V_i|}{\Delta V_i}\right)^2,$$  
where $n=6$ is the number of considered degrees of freedom.  
For the conservative constraints we call for $\chi^2/{\text{d.o.f.}}<2$ and for the aggressive   
ones for $\chi^2/{\text{d.o.f.}}<0.5.$ The choice for the aggressive bounds has been inspired by   
the fact, that one obtains for the best CKM3 fit given by the PDG $\chi^2/{\text{d.o.f.}}=0.4.$ 
In other words, with our aggressive constraints on the tree level bounds, we do not want to    
violate the tree level constraints significantly more than the CKM3 fit.   
From the tree level constraints and careful checks with larger parameter ranges,   
we find that we  safely  restrict ourselves to the ranges given   
in Table \ref{tab:ttl}. The phases $\delta_{13},\delta_{14}$ and $\delta_{24}$ have been   
left unconstrained. The mass $m_{t'}$ was scanned from 300 to 650 GeV as described  
in Equation \eqref{eq:masstp}.    
In this ten-dimensional space we generate more than $2 \cdot 10^{10}$   
randomly distributed points and check whether they meet the tree   
level and FCNC constraints given above.\footnote{A similar strategy with 60 000 points was pursued in \cite{Yanir:2002cq}.}
To this end, we first employ the conservative  
set of bounds. We only store parameter sets which satisfy   
these bounds -- only 12 817 846 data sets remain afterwards.  
The aggressive bounds are established by subsequent reduction of the conservative data, leaving only 150 763 points.   
To give an impression, how important each constraint is under the  
assumption of our preselection, we have used each bound individually and switched off the others.
 
We obtain the following result: Already the tree-level constraints  reduce the allowed parameter space dramatically.  
\begin{table}  
\begin{center}  
\begin{tabular}{|c|c|c|c|c|c|c|}  
 \hline   
&$\Theta_{12}$&$\Theta_{13}$&$\Theta_{23}$&$\Theta_{14}$  
&$\Theta_{24}$&$\Theta_{34}$ \\ \hline \hline  
min. value& $0.222$ & $0.0033$ & $0.038$ & $0$ & $0$ & $0$ \\ \hline  
max. value& $0.232$ & $0.0048$ & $0.046$ & $0.069$ & $0.19$ & $0.8$ \\ \hline  
\end{tabular}  
\caption{Preselection bounds resulting from tree level determinations 
         of the CKM elements for the angles of the quark mixing matrix. \label{tab:ttl}}  
\end{center}  
\end{table}  
Only $13\%$ of the randomly created points in the preselected parameter space actually  
pass the combined tree-level bounds.  The strongest restrictions stem from $|V_{ud}|$,  
which is constrained to a relative error of only 0.028 \%. As a consequence,  
due to $V_{ud}=\cez \ced \cev$, the allowed ranges for $\theta_{12}$ and 
$\theta_{14}$ are quite small ($\theta_{13}$ is tiny, its precise value does not play  
a major role for $|V_{ud}|$). 
Another important contribution to the rejection rate stems from the $\chi^2$ bound.  
The FCNC constraints are even more restrictive, e.g. even in the conservative   
case only $1.5\%$ of the configurations pass the $\Delta_{B_d}$ bound,   
see Table \ref{AcProb:table} for more details.  
\begin{table}  
\center{  
 \begin{tabular}{|c|c|c|c|c|c|}  
  \hline   
&$\Delta_{K^0}$&$\Delta_{B_d}$&$\Delta_{B_s}$&$\Delta_{b\to s\gamma}$  
&$D^0$ mixing \\ \hline \hline  
w/o tree-level bounds&$21\%$ & $1.5\%$ & $29\%$ & $16\%$ & $46\%$ \\ \hline  
w tree-level bounds&$27\%$ & $2.1\%$ & $32\%$ & $20\%$ & $62\%$\\  
\hline  
 \end{tabular}  
\caption{The impact of the (conservative) constraints on the five flavor changing  neutral currents.   
The second line gives the probability that a random point in the configuration space fulfills the FCNC bounds.   
The third line corresponds to the probability that a set of angles and phases that is in agreement with tree-level   
bounds also passes the FCNC bound.\label{AcProb:table}}}  
\end{table}  
Having done our scan, we have found no accepted parameter sets beyond the following ranges:
\begin{displaymath}   
\begin{array}{|l|c|c|}   
\hline   
                   &   \mbox{Conservative Bound} & \mbox{Aggressive Bound}   
\\   
\hline \hline   
\theta_{14}        &    \leq   0.0535             & \leq   0.0364    
\\   
\hline   
\theta_{24}        &    \leq   0.144             & \leq   0.104   
\\   
\hline   
\theta_{34}        &    \leq   0.737             & \leq   0.736    
\\   
\hline   
\delta_{14}        &    \mbox{free}             &  \mbox{free}  
\\   
\hline   
\delta_{24}        &     \mbox{free}             &  \mbox{free}  
\\   
\hline   
\end{array}  
\end{displaymath}  
This  is one of the main results of this work. Typically small mixing
with the fourth family is favoured, but there is still room
for sizeable effects.
To further explain our results, we note that not all combinations for these new parameters are allowed.  
Apart from studying the allowed parameter regions in a one dimensional projection as presented above, we show correlations of selected input parameter pairs.  
Figures \ref{fig:hist3x5}, \ref{fig:hist5x6}, \ref{fig:hist7x8}, \ref{fig:hist8x9} and \ref{fig:hist1x3} correspond to  
the $\theta_{14}-\theta_{24}$, $\theta_{24}-\theta_{34}$, $\delta_{13}-\delta_{14}$,   
$\delta_{14}-\delta_{24}$ and $\theta_{12}-\theta_{14}$  planes. We divide each direction (i.e. x-axis and y-axis) of each plot in 300 steps. So that the  
total picture consists of $300 \times 300=90000$ colour encoded unit squares. In the upper panels the  
colour encoding counts the number of accepted sets in each unit square. As a large range is covered,  
we chose to plot Figures \ref{fig:hist3x5}, \ref{fig:hist5x6}, \ref{fig:hist7x8} and \ref{fig:hist1x3} logarithmically.  
The number next to the colour scale then gives the natural logarithm of the number of accepted sets  
per unit square.  As the distribution in the  $\delta_{14}-\delta_{24}$ plane is somewhat more  
homogeneous we choose a linear scale for Fig.~\ref{fig:hist8x9}.   
The upper left panel in each plot is for the conservative bounds and the upper  
right one for the aggressive ones.  
In the lower panels we present the mass dependence of the allowed parameter ranges. Obviously,  
there is a  non-trivial influence of the $t'$ mass on these ranges. The left panel corresponds to the conservative and the right panel to the 
aggressive bounds. The plots show the  
distribution of the accepted points in the three mass regions indicated in the plot. In most cases
a lower mass results in a larger allowed parameter space. But there are also non-trivial exceptions, cf.  
Figure \ref{fig:hist5x6}. Especially the restriction due to the $D^0$ mixing bound (as described in  
Section 2.2) 
can be seen clearly as hyperbolic cuts in Figure \ref{fig:hist3x5}. 
The mass dependence in Figure \ref{fig:hist8x9} is not shown as in each case  
the whole square is filled.  
\\  
In Figures \ref{fig:K0}, \ref{fig:Bd} and \ref{fig:Bs}, the distribution of the accepted points in the  
complex $\Delta_{[K^0, B_d,B_s]}$ plane is shown. As above, in each plot the left panel corresponds to the conservative  
constraints and the right panel to the aggressive ones. For $\Delta_{K^0}$  
and $\Delta_{B_s}$ a logarithmic scale is chosen and for $\Delta_{B_d}$ a linear one, corresponding  
to the observation that for $B_d$ the points are somewhat more homogeneously distributed as in the other two cases.  
This corresponds to the observation that  
the acceptance rate of the $B_d$ bound is very low, only 2.1 \% after tree level bounds, as shown in Table \ref{AcProb:table}.  
The reason for this behaviour is the following: Enforcing only the tree-level bounds and unitarity $\Delta_{B_d}$ can take values up to $50$ times the Standard Model 
prediction. Therefore, the stringent experimental bounds on $\Delta_{B_d}$
put forward severe restrictions on the allowed parameter range. 

In Fig.~\ref{fig:mdep} the dependence on the $t'$ mass for the three FCNC observables is shown.  
Only for $\Delta_{B_s}$, a strong influence of the mass on the  
results is seen. For $\Delta_{K^0}$ the influence is still perceivable but rather weak,  
whereas $\Delta_{B_d}$ seems to be almost independent of $m_{t'}$.  
The complex $\Delta_B$ planes are particularly interesting since there might be some  
hints on new physics effects in $B_s$ mixing, see \cite{Lenz:2006hd,Tarantino:2009sx}  
and the web-updates of \cite{Hocker:2001xe}.  
In \cite{Lenz:2006hd} a visualization of the combination of the mixing quantities $\Delta M_s$,  
$\Delta \Gamma_s$, $a_{sl}^s$,   which are known to NLO-QCD  
\cite{Buras:1990fn,Beneke:1998sy,Beneke:2003az,Ciuchini:2003ww}   
and of direct determinations of $\Phi_s$ in the complex $\Delta$-plane was suggested.  
Combining recent measurements \cite{HFAG, Tevatron} for the phase $\Phi_s$  
one obtains a deviation from the tiny SM-prediction \cite{Lenz:2006hd} in the range  
of 2 to 3 $\sigma$: 
\begin{itemize} 
\item  HFAG:      2.2 $\sigma$ \cite{HFAG},  
\item  CKM-Fitter: 2.1...2.5 $\sigma$ \cite{Deschamps:2008de, Charles},  
\item  UT-Fit:     2.9 $\sigma$ \cite{Tarantino:2009sx}.  
\end{itemize} 
The central values of these deviations cluster around 
\begin{equation}  
\Phi_s \approx - 45^\circ.  
\end{equation}  
As can be read off from Figure \ref{fig:Bs} sizeable values for $\Phi_s$ can also be obtained  
in scenarios with additional fermions.  
Such large values for $\Phi_s$ are not favoured, but they are possible. An 
enhancement of $\Phi_s$ to large negative values by contributions of a fourth family  
was first discussed in \cite{Hou:2006mx}.  
\\ 
In Figures \ref{fig:Vtb}, \ref{fig:Vts} and \ref{fig:Vtd} we present the  values  
for the CKM matrix elements $V_{tb},V_{ts}$ and $V_{td}$ in the complex plane. As in the Figures \ref{fig:K0}, \ref{fig:Bd}  
and \ref{fig:Bs} the left panel is for the conservative case and the right panel for the aggressive one.  
For comparison the SM3 expectations are given as thin red lines. Obviously, large deviations from the 
SM expectations are possible. The peculiar structure of the allowed range for $V_{td}$ arises already
after imposing unitarity and tree-level constraints. 
The non-trivial  mass dependence for the aggressive case is shown in Fig.~\ref{fig:MassVckm}. 
In Figure \ref{fig:mdist} we show the mass dependence of the acceptance rate. The number of  
accepted data points per 50 GeV normalised to the total number of accepted points is plotted versus   
the mass $m_{t'}$. It can be seen that the acceptance rate reduces with growing $t'$ mass. Because  
our test points are randomly distributed over the whole mass region, an acceptance rate  independent  
from the mass would feature a constant functional behaviour; this is clearly not observed. One can  
also notice a small difference in the acceptance rate for conservative and   
aggressive bounds.
\section{Taylor expansion of $V_{CKM4}$}   
\setcounter{equation}{0}   
The hierarchy of the mixing between the three quark families can be visualized   
by the  Wolfenstein parameterization \cite{Wolfenstein:1983yz}. 
It is obtained from the standard parameterization   
by performing a Taylor expansion in the small CKM element $V_{us} \approx 0.2255$.   
Following \cite{Buras:1994ec} we define   
\begin{eqnarray}   
V_{ub} & = &  s_{13} e^{-i \delta_{13}} =: A \lambda^4 (\tilde{\rho} + i \tilde{\eta})   
\\   
V_{us} & = &  s_{12} (1 + {\cal O} (\lambda^8)) =: \lambda   
\\   
V_{cb} & = &  s_{23} (1 + {\cal O} (\lambda^8)) =: A \lambda^2   
\end{eqnarray}   
Note, that due to historical reasons the element $V_{ub}$ is typically defined to be of order   
$\lambda^3$, while it turned out that it is numerically of order $\lambda^4$.   
\begin{equation}   
|V_{ub}| = 0.00393  = 1.51 \lambda^4 = 0.34 \lambda^3   
\end{equation}   
Up to terms of order $\lambda^6$ the Taylor expansion of the CKM matrix assumes the form: 
{\footnotesize   
\begin{equation}   
 V_{CKM3} =    
\left(    
\begin{array}{ccc}    
1-\frac{\lambda^2}{2}-\frac{\lambda^4}{8} - \frac{\lambda^6}{16}   &    
\lambda  &    
A \lambda^4 (\tilde{\rho} - i \tilde{\eta} )   
\\    
  -\lambda +  A^2 \frac{\lambda^5}{2} -  A^2 \lambda^6 (\tilde{\rho} + i \tilde{\eta}) \, \, \,\, \, \,&     
1-\frac{\lambda^2}{2}-\frac{\lambda^4}{8} -\frac{A^2 \lambda^4}{2}+\frac{A^2 \lambda^6}{4}- \frac{\lambda^6}{16}  \, \, \, \, \, \,&    
A \lambda^2     
\\   
A \lambda^3 - A \lambda^4 (\tilde{\bar{\rho}} + i \tilde{\bar{\eta}}) &    
-A \lambda^2 (1-\frac{\lambda^2}{2} +\lambda^3 (\tilde{\rho} + i \tilde{\eta}) -\frac{\lambda^4}{8} ) &    
1-\frac{A^2 \lambda^4}{2}    
\end{array}    
\right)   
                 .  \end{equation}}
This result can be obtained from the standard Wolfenstein parameterization by replacing  
\begin{equation}  
\rho =: \lambda \tilde{\rho}\, , \hspace{1cm} \eta =: \lambda \tilde{\eta} \, .  
\end{equation}  
For the case of 4 generations we have to determine first the possible size, i.e. the power in $\lambda$   
of the new CKM-matrix elements. With the results of the previous section we obtain:   
\begin{displaymath}   
\begin{array}{|l|c|c|}   
\hline   
                   &   \mbox{Conservative Bound} & \mbox{Aggressive Bound}   
\\   
\hline \hline   
|V_{ub'}|        &    \leq   0.0535 \approx 1.05 \lambda^2  & \leq   0.0364 \approx 0.7 \lambda^2 \approx 3.2 \lambda^3    
\\   
\hline   
|V_{cb'}|        &    \leq   0.144  \approx 0.6 \lambda^1 \approx 2.8 \lambda^2  & \leq   0.104 \approx 0.46 \lambda^1 \approx 2 \lambda^2    
\\   
\hline   
|V_{tb'} |       &    \leq   0.672  \approx 3.0 \lambda^1  & \leq   0.671 \approx 3.0 \lambda^1    
\\   
\hline   
\end{array}   
\end{displaymath} 
We propose a parameterization of these matrix elements 
that manifestly respects the above bounds:   
%
%
%
\begin{itemize}   
\item For the mixing of first and fourth family we define  
      \begin{eqnarray}   
      V_{ub'} & = & s_{14} e^{-i \delta_{14}} =: \lambda^2 (x_{14} -i y_{14} ) \nn    
      \\   
        & \Rightarrow & s_{14} = \lambda^2 \sqrt{x_{14}^2 + y_{14}^2} \nn   
      \\   
        & \Rightarrow & c_{14} = 1 - \lambda^4 \frac{x_{14}^2 + y_{14}^2}{2}+O\left(\lambda ^8\right)\; ,  
      \end{eqnarray}   
      which is a good estimate for both, conservative and aggressive bounds, since the  
      parameters $x_{14}$ and $y_{14}$ can safely be assumed to be smaller than $1$.  
\item The estimate for the matrix element $V_{cb'}$ is more complicated. The conservative   
      bound suggests a size of order $\lambda$, whereas the aggressive bound might justify  
      a leading power $\lambda^2$.  In what follows we opt for the more solid $\mathcal{O}(\lambda)$  
      variant. We define:  
      \begin{align} 
      V_{cb'}&= c_{14} s_{24} e^{-i \delta_{24}}=: (x_{24} - iy_{24}) \lambda^1 \nn \\ 
      & \Rightarrow  s_{24}  e^{-i \delta_{24}} =\left(x_{24}-i y_{24}\right) \lambda +\frac{1}{2} \left(x_{14}^2+y_{14}^2\right) \left(x_{24}-i y_{24}\right) \lambda ^5+O\left(\lambda ^7\right)
\nn \\ 
&\Rightarrow c_{24}=
\label{VcbpIsConfusing} 
1+\frac{1}{2} \left(-x_{24}^2-y_{24}^2\right) \lambda ^2-\frac{1}{8} \left(x_{24}^2+y_{24}^2\right){}^2 \lambda ^4 
\nn \\
& \qquad \qquad \qquad +\frac{1}{6} \left(\frac{3}{8}
   \left(-x_{24}^2-y_{24}^2\right){}^3+3 \left(-x_{14}^2-y_{14}^2\right) \left(x_{24}^2+y_{24}^2\right)\right) \lambda ^6+O\left(\lambda ^7\right)
\end{align} 
 \item Finally, the element $|V_{tb'}|$ is not constrained to be significantly smaller than one 
	and we cannot restrict the mixing angle $\Theta_{34}$. Thus, we keep cosine $c_{34}$ and 
	sine $s_{34}$ in the expansion. 
\end{itemize} 

It is obvious that already at $\mathcal{O}(\lambda^6)$ the expansion gets confusing, 
see \eqref{VcbpIsConfusing}. For the Taylor expansion to provide an intuitive  
picture of the hierarchy of the elements and the still possible effects of the mixing  
with the fourth generation we want to keep the matrix  clearly arranged. 
Therefore we expand the CKM4 matrix up to and including order $\lambda^4$. 
The matrix elements take the form
\begin{align}
 V_{ud}&=1-\frac{\lambda ^2}{2}-\frac{1}{8} \left(4 x_{14}^2+4 y_{14}^2+1\right) \lambda ^4 & V_{us}&=\lambda   
\nn \\
 V_{ub}&= A ({\tilde{\rho}} -i {\tilde{\eta}}) \lambda ^4 & V_{ub'}&= \left(x_{14}-i y_{14}\right) \lambda ^2 \nn \\
\end{align}
\begin{align}
V_{cd}=&-\lambda +\frac{1}{2} \left(x_{24}-i y_{24}\right) \left(-2 x_{14}+x_{24}-2 i y_{14}+i y_{24}\right) \lambda ^3
\nn \\ 
V_{cs}= &1-\frac{1}{2} \left(x_{24}^2+y_{24}^2+1\right) \lambda ^2 \nn \\
&\phantom{1} +\frac{1}{8} \left(-x_{24}^4-2 \left(y_{24}^2-1\right) x_{24}^2-8 i y_{14} x_{24}-y_{24}^4 \right. 
\nn \\
&\qquad\qquad\left.-4 A^2+2
   y_{24}^2-8 x_{14} \left(x_{24}-i y_{24}\right)-8 y_{14} y_{24}-1\right) \lambda ^4
\nn \\ 
V_{cb}= & A \lambda ^2 \nn \\
V_{cb'}=& \left(x_{24}-i y_{24}\right)\lambda  
\\
V_{td}= &\textcolor{WildStrawberry}{s_{34} \left(-x_{14}+x_{24}-i \left(y_{14}-y_{24}\right)\right) \lambda ^2} \nn \\
& +A c_{34} \lambda ^3 +\frac{1}{2} \left[ A (-2 i {\tilde{\eta}} -2 {\tilde{\rho}} ) c_{34}+s_{34} 
   \left(x_{14}+i y_{14}\right) \left(x_{24}^2+y_{24}^2+1\right)\right] \lambda ^4 \nn \\
V_{ts}=&\textcolor{WildStrawberry}{-s_{34} \left(x_{24}+i y_{24}\right) \lambda} -A c_{34} \lambda ^2 \nn \\
&\quad+\frac{1}{2} s_{34} \left(-2 x_{14}+x_{24}-2 i y_{14}+i y_{24}\right) \lambda ^3-\frac{1}{2}
   \left[A c_{34} \left(x_{24}^2+y_{24}^2-1\right)\right] \lambda ^4 \nn \\
V_{tb}=& c_{34}-A s_{34} \left(x_{24}+i y_{24}\right) \lambda ^3 -\frac{1}{2} \left(A^2 c_{34}\right) \lambda ^4 \nn \\
V_{tb'}=&s_{34}-\frac{1}{2} \left[s_{34} \left(x_{24}^2+y_{24}^2\right)\right] \lambda ^2 \nn \\
&\qquad-\frac{1}{8} \left[s_{34} \left(x_{24}^4+2 y_{24}^2 
   x_{24}^2+y_{24}^4+4 x_{14}^2+4 y_{14}^2\right)\right] \lambda ^4 
\\
 V_{t'd}=& c_{34} \left[-x_{14}+x_{24}-i \left(y_{14}-y_{24}\right)\right] \lambda ^2  
\nn\\
&\qquad-A s_{34} \lambda ^3+\frac{1}{2} \left[2 A (i {\tilde{\eta}} +{\tilde{\rho}} ) s_{34}+c_{34} \left(x_{14}+i 
   y_{14}\right) \left(x_{24}^2+y_{24}^2+1\right)\right] \lambda ^4 \nn \\
V_{t's}=& -c_{34} \left(x_{24}+i y_{24}\right) \lambda +A s_{34} \lambda^2 \nn \\
&\qquad+\frac{1}{2} c_{34} \left(-2 x_{14}+x_{24}-2 i y_{14}+i y_{24}\right) \lambda ^3 +\frac{1}{2} A s_{34} \left(x_{24}^2+y_{24}^2-1\right) \lambda 
   ^4 \nn \\
V_{t'b}=&-s_{34}-A c_{34} \left(x_{24}+i y_{24}\right) \lambda ^3 + \frac{1}{2} A^2 s_{34} \lambda ^4  \nn \\
V_{t'b'}=&c_{34}-\frac{1}{2} \left[c_{34} \left(x_{24}^2+y_{24}^2\right)\right] \lambda ^2 
\nn \\ &\qquad-\frac{1}{8} \left[c_{34} \left(x_{24}^4+2 y_{24}^2 
   x_{24}^2+y_{24}^4+4 x_{14}^2+4 y_{14}^2\right)\right] \lambda ^4 
\end{align}
The \textcolor{WildStrawberry}{red} colored terms indicate possible new {\it leading order} effects in the standard CKM3 matrix elements due to mixing with the fourth family. 
\section{Unexpected parameter regions}   
 \setcounter{equation}{0}  
In the experimentally allowed regions of the parameter space we  typically find regions, where the   
mixing with the fourth family is very small and the CKM elements of the first three families  
are close to the minimal standard model values.  
There are also some allowed regions with large deviations from the standard expectations.  
In order to clarify the appearing cancellations, that veil these unexpected effects 
in current analyses of the standard CKM matrix, we discuss three  
sample sets of values for $V_{CKM4}$. 
Our three parameter sets read: 
\begin{displaymath} 
\begin{array}{|c|c|c|c|} 
\hline  
            & \mbox{Set I} & \mbox{Set II} & \mbox{Set III} 
\\ 
\hline \hline 
\theta_{12} & 0.226606 & 0.227264   & 0.228225 
\\  
\hline 
\theta_{23} & 0.040389 &0.0414083   & 0.039522   
\\  
\hline 
\theta_{13} & 0.0040559 & 0.00382191& 0.00382755  
\\  
\hline 
\theta_{14} & 0.0277527 & 0.0182248 & 0.0232895  
\\  
\hline 
\theta_{24} & 0.0176553 & 0.0789555 & 0.110918  
\\  
\hline 
\theta_{34} &-0.531735 & 0.366353   & 0.677976  
\\  
\hline 
\delta_{13} & 3.31463  & 0.317332   & 1.25537  
\\  
\hline 
\delta_{14} & 0.925439 & 0.28357    & 0.502528 
\\  
\hline 
\delta_{24} & 2.69829 & 0.383156    & 0.238529 
\\  
\hline 
m_{t'}      & 325.553 \, \mbox{GeV} & 653.842 \, \mbox{GeV} & 389.238 \, \mbox{GeV}  
\\ 
\hline 
\end{array} 
\end{displaymath} 
First we have a look at the CKM elements $V_{tx}$ obtained with these three parameter sets. 
We give their complex values, as well as the ratio of their absolute value compared to the SM3 values from 
\cite{Amsler:2008zzb}: 
\begin{displaymath} 
\begin{array}{|c|c|c|c|} 
\hline 
                        & \mbox{Set I} & \mbox{Set II} & \mbox{Set III} 
\\ 
\hline \hline 
V_{td}                  &  0.0212 + 0.0107 i  & 0.0052 - 0.0005 i   & 0.0089 - 0.0059 i 
\\  
\hline 
|V_{td}|/|V_{td}^{SM3}| &   2.72              &  0.60               &  1.22 
\\  
\hline 
V_{ts}                  &  -0.0391 + 0.0064 i & -0.0653 - 0.0109 i   & -0.0987 - 0.0182 i 
\\  
\hline 
|V_{ts}|/|V_{ts}^{SM3}| &   0.97              &  1.63               &  2.47 
\\  
\hline 
V_{tb}                  &  0.8609 + 0.0001 i  & 0.9317 - 0.0004 i   & 0.7755 - 0.0006 i 
\\ 
\hline 
|V_{tb}|/|V_{tb}^{SM3}| &   0.86              &  0.93               & 0.78 
\\  
\hline 
\end{array} 
\end{displaymath}  
These results significantly differ from the values obtained from SM3 CKM fits.  
In order to clarify the question why these huge effect cannot be seen in  
the standard CKM-fits \cite{Hocker:2001xe,Tarantino:2009sx} 
we have a closer look at e.g. $\Delta_{B_d}$.  
This quantity was defined as 
\begin{eqnarray}  
\Delta_{B_d} & = & \frac{M_{12, SM4}^{B_d}}{M_{12, SM3}^{B_d}}  
= \frac{M_{12, SM4}^{tt,B_d} + M_{12, SM4}^{(tt'+t't'),B_d}}{M_{12, SM3}^{B_d}} \, . 
\end{eqnarray}  
The $tt$ part of the SM4 value $M_{12, SM4}^{tt,B_d}$ looks formally equal to  
$M_{12, SM3}^{B_d}$, but the values of the CKM elements $V_{tx}$ can be very different   
for SM3 and SM4.  
We further rewrite $\Delta_{B_d}$ as  
\begin{eqnarray}  
\Delta_{B_d} & = & 1 +   
\frac{M_{12, SM4}^{tt,B_d} - M_{12, SM3}^{B_d}}{M_{12, SM3}^{B_d}}  
+ \frac{M_{12, SM4}^{tt'+t't',B_d}}{M_{12, SM3}^{B_d}} \, . 
\end{eqnarray}  
The first correction term to ``1'' is due to the difference of the CKM elements $V_{tx}$ in the three 
and four generation standard model, while the second correction is due to new virtual loop effects of the  
$t'$ quark. The three parameter sets, discussed in this section, were chosen in such a way that large  
cancellations appear mimicking the SM3 perfectly. Therefore, these big effects  
are invisible in CKM fits. 
\\ 
With our special parameter sets we numerically obtain the following
values for the three contributions to $\Delta$:  
\\ 
\centerline{Set I:} 
\begin{eqnarray}  
\Delta_{K^0} & = & 1 +    (0.0139 -0.0854 i )  + (-0.0362 + 0.0416 i) 
\nonumber \\ 
             & = & 0.98 \cdot e^{-i2.5^\circ},   
\\ 
\Delta_{B_d} & = & 1 +    (-1.6939 -5.4548 i )  + (1.7352 + 5.3184 i )  
\nonumber \\ 
             & = & 1.05 \cdot e^{-i7.5^\circ},   
\\ 
\Delta_{B_s} & = & 1 +    (-0.3415 + 0.2492 i)  + (0.3608 - 0.3662 i)  
\nonumber \\ 
             & = & 1.03 \cdot e^{-i6.5^\circ},   
\\ 
\Delta_{b \to s \gamma} & = & 1 -0.2959  + 0.3715 = 1.0756 \,,
\\ 
\Phi^\Delta_s &  = & -0.114276 = -6.5^\circ \,.
\end{eqnarray}  
Huge cancellations appear, in the case of the imaginary part of $B_s$ mixing up to 500\%.
Taking experimental and theoretical uncertainties into account, the final results are still
perfectly consistent with the SM3 expectation.
\\ 
For the next parameter set we get:
  
\centerline{Set II:} 
\begin{eqnarray}  
\Delta_{K^0} & = & 1 +    (-0.0016 -0.0017 i )  + (-0.0246 - 0.0071 i) 
\nonumber \\ 
             & = & 0.97 \cdot e^{-i 0.5^\circ},   
\\ 
\Delta_{B_d} & = & 1 +    (-0.7383 - 0.1732 i )  + (0.4631 + 0.0826 i )  
\nonumber \\ 
             & = & 0.73 \cdot e^{-i 7.1^\circ},   
\\ 
\Delta_{B_s} & = & 1 +    (1.2044 - 0.6715 i)  + (-1.3434 - 0.0354 i)  
\nonumber \\ 
             & = & 1.11 \cdot e^{-i 39^\circ},   
\\ 
\Delta_{b \to s \gamma} & = & 1 + 1.3044  - 1.3879 = 0.9165 \,,
\\ 
\Phi^\Delta_s &  = & -0.687 = -39^\circ \,.
\end{eqnarray}  
This set was chosen by looking for large values of $\Phi_s$. As discussed in Section 2.3 
there are currently some experimental hints for such a deviation from the standard model. 
Here we confirm the statement from \cite{Hou:2006mx} that such a value could be explained 
by a forth generation of quarks. 
\\ 
As a final example we present a parameter set yielding a value for $|V_{tb}|$ 
as small as $0.78$.
 
\centerline{Set III:} 
\begin{eqnarray}  
\Delta_{K^0} & = & 1 +    (0.0108 +0.0919 i )  + (-0.0388 - 0.0106 i) 
\nonumber \\ 
             & = & 0.98 \cdot e^{+i 4.8^\circ},   
\\ 
\Delta_{B_d} & = & 1 +    (-0.1691 + 0.3448 i )  + (0.1681 - 0.4824 i )  
\nonumber \\ 
             & = & 1.01 \cdot e^{-i 7.8^\circ},   
\\ 
\Delta_{B_s} & = & 1 +    (2.4697 - 1.1837 i)  + (-2.8227 + 0.8334 i)  
\nonumber \\ 
             & = & 0.74 \cdot e^{-i 28^\circ},   
\\ 
\Delta_{b \to s \gamma} & = & 1 + 2.6661  - 2.7172 = 0.9489 \,,
\\
\Phi^\Delta_s &  = & -0.4961 = -28^\circ \,.
\end{eqnarray}  
A small value of $V_{tb}$ would also lead to a smaller rate for e.g. the single top production 
at TeVatron. See e.g.  \cite{st:2008nf} for a recent measurement of this rate. 

Note that the effects described in the chosen sets are very sensitive to 
small variations in the mixing angles and phases of the fourth family. This is obvious as the large cancellations described above require 
very specific parameter sets. The dependence on the $t'$ mass, in contrast, is moderate.
   
\section{Conclusion}   
\setcounter{equation}{0}

We have investigated the experimentally allowed parameter range  
for a 4$\times$4 quark mixing matrix, making some simplifying assumptions  
concerning the QCD corrections. Moreover we have not taken into account  
any correlations with the lepton mixing matrix.  
\\  
As a result we find that the tree-level constraints for the 3$\times$3 CKM-matrix and the FCNC bounds  
from $K$-, $D^0$- $B_d$- and $B_s$-mixing as well as the decay $b \to s \gamma$ are typically fulfilled   
if we have a small mixing with the fourth family, which allows us to perform 
a Taylor expansion of the 4$\times$4 CKM matrix.
Unexpectedly we were also able to find experimentally allowed parameter sets, having a sizeable mixing with the   
fourth generation.  
In this case also the usual 3$\times$3 CKM matrix elements can change considerably:  
$V_{td}$ and $V_{ts}$ can differ by up to a factor $3$ compared to the SM3 value and   
 $V_{tb}$ can be as low as 0.75,   
see also \cite{Alwall:2006bx} for the possibility of $V_{tb}$ being unequal to one.  
These dramatic effects are not seen in the CKM fits. This is due to large cancellations between 
the effect of changed matrix elements $V_{tx}$ and effects of virtual heavy $b'$ and $t'$ quarks.   
An example of such a cancellation was also discussed in \cite{Hou:2005yb}.
We have also shown that there are parameter ranges consistent with all experimental 
bounds, which yield large effects for $\Phi_s$.  
\\  
Due to these interesting results, it seems worthwhile to extend the current exploratory analysis.  
First more flavor observables, like asymmetries, $b \to s l^+ l^-$ (see e.g. 
\cite{Aliev:2000cm,Buchalla:2000sk}),
$B_s \to \mu \mu$, ... should be considered.  
Moreover, the electro-weak precision observables have to be included in more detail, here in particular
the observable $R_b$ seems to be promising, see e.g \cite{Bamert:1996px}.
Another important improvement will be the exact treatment of the perturbative QCD corrections,   
in particular in the decay $b \to s \gamma$. Finally one has also to take into account correlations to   
the lepton mixing matrix.  
\\  

Refined direct measurements of the CKM matrix elements will provide 
more insight into a possible fourth generation. In particular, future experiments could help 
to determine the hardly known CKM elements $V_{cd}$ and $V_{cs}$ as well as non-perturbative parameters
like form factors and decay constants.
Probably the most stringent bounds on the mixing with the fourth generation can be obtained from  
the direct measurements of $V_{td}$, $V_{ts}$ and $V_{tb}$. $V_{tb}$ is currently investigated  
at TeVatron; for the latest  value of $V_{tb}$ from  single top production see 
\cite{st:2008nf,Abazov:2009ii,Aaltonen:2009jj}.  
\\  
Also more precise data on FCNC will be very helpful. For the case of the promising $B_s$ system this is   
currently done at the TeVatron and in the near future at LHCb \cite{LHCb} and probably at Super B factories  
\cite{Bona:2007qt,Hashimoto:2004sm}.

\section*{Acknowledgements}   
    
We thank Ayres Freitas, Bob Holdom, George Hou, Emi Kou, Graham Kribs, Heiko Lacker, 
Michael Spannowsky and Wolfgang Wagner for clarifying discussions.  
M.~B. is supported by BayBFG and Studienstiftung des deutschen Volkes, Bonn, Germany.
J.~Riedl is supported by a grant of the Cusanuswerk, Bonn, Germany.

\end{document}